%% file: main.tex
\pdfoutput=1 
\documentclass[a4paper,10pt]{article}
\usepackage{geometry}
\usepackage{graphics}  
\usepackage{graphicx}
\usepackage{subfigure}
\usepackage{amsmath}
\usepackage[amssymb]{SIunits}
\usepackage{nicefrac}
\usepackage[english]{babel}
\usepackage{lineno}
\usepackage{epstopdf}
\usepackage{stfloats}
\usepackage{verbatim}
\usepackage{url}
\usepackage{xcolor}
\usepackage{url}
\usepackage{setspace}

\newcommand{\DESY}{\ensuremath{\textrm{DESY}}}
\newcommand{\Datura}{\ensuremath{\textrm{DATURA}}}

\newcommand{\Mimosa}{\ensuremath{\textrm{MIMOSA\,26}}}
\newcommand{\noise}{\ensuremath{\xi_{\textrm{n}}}}
\newcommand{\epsdut}{\ensuremath{\mathnormal{\varepsilon_{\textrm{DUT}}}}}
\newcommand{\epsmimo}{\ensuremath{\mathnormal{\varepsilon_{\textrm{M26}}}}}
\newcommand{\dz}{\ensuremath{\textrm{d}z}}
\newcommand{\dzdut}{\ensuremath{\textrm{d}z_{\textrm{DUT}}}}

\newcommand{\sigmatu}{\ensuremath{\sigma_{\textrm{t,u}}}}
\newcommand{\sigmatb}{\ensuremath{\sigma_{\textrm{t,b}}}}
\newcommand{\sigmat}{\ensuremath{\sigma_{\textrm{t}}}}

\newcommand{\sigmai}{\ensuremath{\sigma_{\textrm{int}}}}
\newcommand{\sigmam}{\ensuremath{\sigma_{\textrm{M26}}}}
\newcommand{\sigmahat}{\ensuremath{\hat{\sigma}_{\textrm{int}}}}

\newcommand{\zdut}{\ensuremath{z_{\textrm{DUT}}}}

\newcommand{\rbiased}{\ensuremath{r_{\textrm{b}}}}
\newcommand{\runbiased}{\ensuremath{r_{\textrm{u}}}}

\newcommand{\pb}{\ensuremath{p_{\textrm{b}}}}

\newcommand{\eudet}{\ensuremath{\textrm{EUDET}}}

\newcommand{\eudaq}{\ensuremath{\textrm{EUDAQ}}}
\newcommand{\EUTelescope}{\ensuremath{\textrm{EUTelescope}}}

\newcommand*{\notFOREPJ}{}%

\setpagewiselinenumbers
\modulolinenumbers[5]

\ifdefined\notFOREPJ
\else
\fi

\makeatletter
\renewcommand{\maketitle}{\bgroup\setlength{\parindent}{0pt}
\begin{flushleft}
  \vspace*{10mm}
  \textbf{\huge\sffamily\@title}
  \vspace{5mm}
   
  \large \@author
\end{flushleft}\egroup
}
\def\@xfootnote[#1]{%
  \protected@xdef\@thefnmark{#1}%
  \@footnotemark\@footnotetext}
\makeatother

\begin{document}







\title{Performance of the EUDET-type\\ beam telescopes}
\author{
H.~Jansen${}^{\textrm{a,}}$\footnote[*]{Corresponding author: hendrik.jansen@desy.de},
S.~Spannagel${}^{\textrm{a}}$, 
J.~Behr${}^{\textrm{a,}}$\footnote{Now at Institut f\"ur Unfallanalysen, Hamburg, Germany},
A.~Bulgheroni${}^{\textrm{b,}}$\footnote{Now at KIT, Karlsruhe, Germany},
G.~Claus${}^{\textrm{c}}$,
E.~Corrin${}^{\textrm{d,}}$\footnote{Now at SwiftKey, London, UK},
D.~G.~Cussans${}^{\textrm{e}}$,
J.~Dreyling-Eschweiler${}^{\textrm{a}}$, 
D.~Eckstein${}^{\textrm{a}}$, 
T.~Eichhorn${}^{\textrm{a}}$, 
M.~Goffe${}^{\textrm{c}}$,
I.~M.~Gregor${}^{\textrm{a}}$, 
D.~Haas${}^{\textrm{d,}}$\footnote{Now at SRON, Utrecht, Netherlands},
C.~Muhl${}^{\textrm{a}}$,
H.~Perrey${}^{\textrm{a,}}$\footnote{Now at Lund University, Sweden}, 
R.~Peschke${}^{\textrm{a}}$, 
P.~Roloff${}^{\textrm{a,}}$\footnote{Now at CERN, Geneva, Switzerland}, 
I.~Rubinskiy${}^{\textrm{a,}}$\footnote{Now at CFEL, Hamburg, Germany}, 
M.~Winter${}^{\textrm{c}}$\\
\vspace{3mm}
${}^{\textrm{a}}$ Deutsches Elektronen-Synchrotron DESY, Hamburg, Germany\\
${}^{\textrm{b}}$ INFN Como, Italy\\
${}^{\textrm{c}}$ IPHC, Strasbourg, France\\
${}^{\textrm{d}}$ DPNC, University of Geneva, Switzerland\\
${}^{\textrm{e}}$ University of Bristol, UK
}
\maketitle


\begin{abstract}
\noindent
Test beam measurements at the test beam facilities of DESY have been conducted to characterise the performance of the EUDET-type beam telescopes originally developed within the $\eudet$ project. 
The beam telescopes are equipped with six sensor planes using $\Mimosa$ monolithic active pixel devices. 
A programmable Trigger Logic Unit provides trigger logic and time stamp information on particle passage. 
Both data acquisition framework and offline reconstruction software packages are available. 
User devices are easily integrable into the data acquisition framework via predefined interfaces.
 
The biased residual distribution is studied as a function of the beam energy, plane spacing and sensor threshold. 
Its standard deviation at the two centre pixel planes using all six planes for tracking in a 6\,GeV electron/positron-beam is measured to be $(2.88\,\pm\,0.08)\,\upmu\meter$.
Iterative track fits using the formalism of General Broken Lines are performed to estimate the intrinsic resolution of the individual pixel planes. 
The mean intrinsic resolution over the six sensors used is found to be $(3.24\,\pm\,0.09)\,\upmu\meter$.
With a 5\,GeV electron/positron beam, the track resolution halfway between the two inner pixel planes using an equidistant plane spacing of 20\,mm is estimated to $(1.83\,\pm\,0.03)\,\upmu\meter$
 assuming the measured intrinsic resolution. 
Towards lower beam energies the track resolution deteriorates due to increasing multiple scattering. 
Threshold studies show an optimal working point of the $\Mimosa$ sensors at a sensor threshold of between five and six times their RMS noise. 
Measurements at different plane spacings are used to calibrate the amount of multiple scattering in the material traversed
 and allow for corrections to the predicted angular scattering for electron beams. 
\end{abstract}

\ifdefined\notFOREPJ
\tableofcontents
\else
\fi

\section{Introduction}
\label{sec:intro}
\ifdefined\notFOREPJ
\input{content/intro}
\else
\input{intro}
\fi

\section{Beamlines}
\label{sec:beamlines}
\ifdefined\notFOREPJ
\input{content/beam_lines}

\else
\input{beam_lines}
\fi

\section{Components of the EUDET-type beam telescopes}
\label{sec:tscope}
\ifdefined\notFOREPJ
\input{content/tscope}

\else
\input{tscope}
\fi

\section{The EUDAQ data acquisition framework}
\label{sec:eudaq}
\ifdefined\notFOREPJ
\input{content/eudaq}
\else
\input{eudaq}
\fi

\section{Offline analysis and reconstruction using EUTelescope}
\label{sec:offline}
\ifdefined\notFOREPJ
\input{content/offline}
\else
\input{offline}
\fi

\section{Track resolution studies}
\label{sec:trackres}
\ifdefined\notFOREPJ
\input{content/track_resolution}
\else
\input{track_resolution}
\fi

\section{Considerations for DUT integrations}
\label{sec:dutintegration}
\ifdefined\notFOREPJ
\input{content/dutintegration}
\else
\input{dutintegration}
\fi

\section{Conclusion}
\label{sec:conclusion}
\ifdefined\notFOREPJ
\input{content/conclusion}
\else
\input{conclusion}
\fi

\section*{Data and materials}
The datasets supporting the conclusions of this article are available from reference \cite{jansen_data}.
The software used is available from the github repositories: 1) \url{https://github.com/eutelescope/eutelescope}, 2) \url{https://github.com/simonspa/eutelescope/}, branch \textit{scattering}
 and 3) \url{https://github.com/simonspa/resolution-simulator}.
For the presented analysis, these specific tags have been used: \cite{jansen_2016_49065} and \cite{spannagel_2016_48795}.

\section*{Competing interests}
The authors declare that they have no competing interests.

\section*{Acknowledgements}
We are indebted to Claus Kleinwort for his counsel and numerous discussions. 
Also, we would like to thank Ulrich K\"otz.
The test beam support at DESY is highly appreciated. 
This work was supported by the Commission of the European Communities under the FP7 Structuring the European Research Area, contract number RII3-026126 (EUDET). 
Furthermore, strong support from the Helmholtz Association and the BMBF is acknowledged.

\small
\bibliographystyle{IEEEtran}
\ifdefined\notFOREPJ
\bibliography{bibtex/refs}
\else
\bibliography{refs}
\fi

\end{document}

%% file: content/intro.tex
Beam telescopes are vital tools for R\&D projects focussing on position sensitive particle detection sensors. 
These range from collider-specific detectors with high radiation tolerance~\cite{1748-0221-9-12-C12001,1748-0221-9-12-C12029},
 high resolution and low material requirements~\cite{1748-0221-10-03-C03044} to medical applications~\cite{Ballabriga2011S15}, among others. 
Complementary to sensor simulations using finite element analysis tools, test beam studies are used at various stages of sensor and read-out chip development. 
Such test beam studies are well suited and often used for the evaluation of the performance of a detector prototype. 

Within the Integrated Infrastructure Initiative funded by the EU in the 6th framework programme,
 the EUDET project aimed at providing a high-resolution pixel beam telescope for test beam studies~\cite{ref:eudetreport200902}.
The guidelines for the development were to allow for an easy integration of custom data acquisition systems covering a wide range of readout schemes, latencies, and acquisition rates.
This is achieved by well defined interfaces on both the hardware and the software level. 
Fast LHC-type tracking devices are integrable in the same manner as slower rolling-shutter readout devices. 

The EUDET-type beam telescope each consist of six pixel detector planes equipped with fine-pitch $\Mimosa$ sensors~\cite{HuGuo2010480},
 the mechanics for precise positioning of the device under test (DUT) and the telescope planes in the beam, a Trigger Logic Unit (TLU) providing trigger capabilities and a data acquisition system.
The chosen design meets most user requirements in terms of easy integration capabilities, spatial resolution, and trigger rates. 
The telescope planes are designed and built to keep the material budget as low as possible in order to achieve an excellent track resolution
 even at the rather low particle energies of up to 6\,GeV at the DESY test beam facilities.

The original EUDET beam telescope, which was modified to become the AIDA telescope, is operated at SPS beamline H6 (CERN).
Responding to the increasing demand of the sensor R\&D community, several replicas, collectively called $\eudet$-type beam telescopes, have been built since then:
 ACONITE for the ATLAS group, which is also operated at the beamline H6, ANEMONE at ELSA (University of Bonn), the copy for the Carlton University called CALADIUM, 
 and two copies, $\Datura$ and DURANTA, which are operated at DESY. 
 Within the AIDA2020 project, another copy is going to be built -- operation is foreseen at the PS beamline (CERN).
All replicas are based on the $\Mimosa$ sensors and are equipped with the same data acquisition system and software framework. 
The EUDET telescope has been used since 2007 in various stages of development by hundreds of users and played an important role in sensor studies employed by a wide community. 
From January 2013 until March 2014 alone, about 300 users utilised an $\eudet$-type beam telescope at DESY for a total of 80 user weeks. 
The results reported here are based on data taken with $\Datura$ at test beam area~21 at {DESY-II} and are comparable to other beam telescope copies with
a similar thickness of the epitaxial layer~\cite{desy-tscopes-main}. 

This paper is organised as follows: 
The DESY beamlines are introduced in section~\ref{sec:beamlines}, followed by the description of the beam telescope
 and the data acquisition framework in sections~\ref{sec:tscope} and~\ref{sec:eudaq}, respectively.
Section~\ref{sec:offline} details the offline analysis and reconstruction software. 
Results of the EUDET-type beam telescope performance and track resolution predictions for different telescope configurations and beam momenta are presented in section~\ref{sec:trackres}. 
Additionally, predictions of the standard deviation of the angular scattering distribution are compared to measurements with an electron beam. 
Section~\ref{sec:dutintegration} discusses the integration of DUTs into an $\eudet$-type beam telescope. 

%% file: content/beam_lines.tex
For this work, data have been taken at the {DESY-II} test beam facilities. 
The DESY-II electron/positron synchrotron at the DESY site in Hamburg has a circumference of 292.8\,m and is mainly used as an injector for the PETRA-III storage ring. 
However, it also supplies beam to three test beam areas, one  of them being the test beam area~21.
Its dipole magnets operate in a sinusoidal ramping mode with a frequency of 12.5\,Hz. 
Therefore, one DESY-II cycle takes 80\,ms, and the bunch length is around 30\,ps. 
The DESY-II synchrotron is equipped with movable carbon fibres. 
If positioned in the beam, bremsstrahlung photons are created, escaping the beam line tangentially.
Subsequently, the photons are converted to electron/positron pairs on a secondary metal target. 
Their energy distribution reaches up to 6\,GeV. 
Using a dipole magnet, this secondary electron/positron beam is spread out.
A collimator can be used to select certain energy ranges of the beams reaching the experimental halls with an achievable rate of about $10$\,kHz to $100$\,kHz. 
A more detailed description of the test beams at {DESY-II} can be found in~\cite{EUDET-2007-11}.





%% file: content/tscope.tex
\begin{figure}[tb]
	\center
	\ifdefined\notFOREPJ
	\includegraphics[width=.9\textwidth]{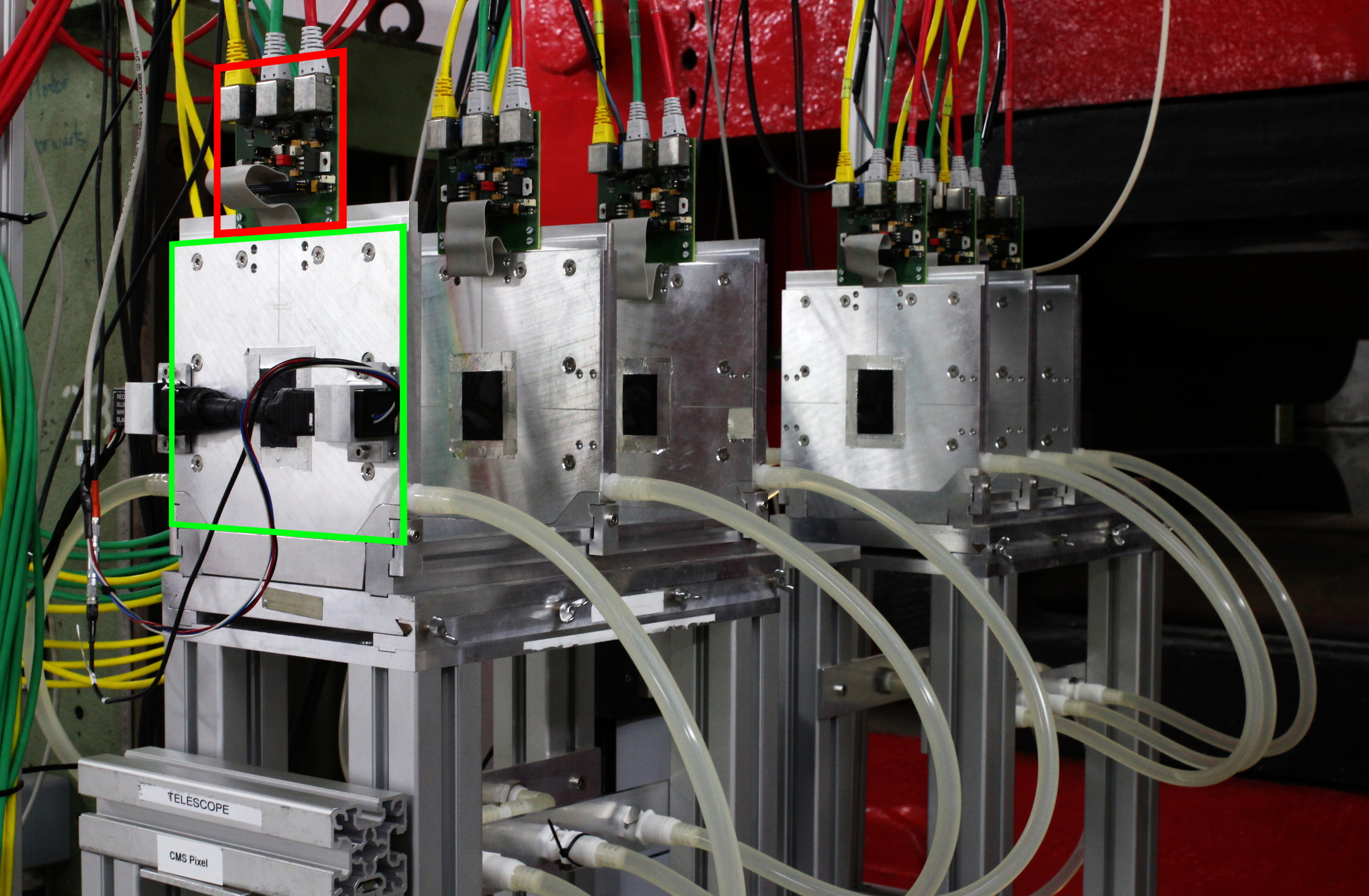}
	\else
	\includegraphics[width=.9\textwidth]{DATURA2a.jpg}
	\fi
	\caption[The $\Datura$ telescope]{The $\Datura$ beam telescope with its sensor planes (green frame) on aluminium rails.
	The auxiliary boards (red frame) with connections for clock, sensor configuration, and data are mounted at the top of the planes.
	Coolant is provided to the planes via tubes.}
	\label{fig:datura-tscope}
\end{figure}

The EUDET-type beam telescopes are tabletop tracking detectors featuring six pixelated silicon sensors, four scintillators with photo multiplier tubes (PMTs) for trigger purposes,
 a Trigger Logic Unit (TLU) providing trigger logic and time stamp information on particle passage, and a data acquisition system for the readout. 
Figure\,\ref{fig:datura-tscope} shows the beam telescope with its rectangular aluminium sensor jigs, wherein the pixel sensors are embedded,
 and its auxiliary boards providing connections for power, sensor configuration, clock signals and data transmission. 
The planes are organised in two telescope arms holding three sensors each. 
A DUT can be inserted between the arms or at either end of the beam telescope. 
In the Cartesian, right-handed coordinate system chosen, the $y$-direction points vertically down and the $z$-direction along the beam direction.

\subsection{Sensors and mechanics}
\label{sec:sensors}

The $\Mimosa$ sensors used for precise spatial measurements of particle trajectories are manufactured with the AMS 350\,nm CMOS technology~\cite{HuGuo2010480}. 
Each $\Mimosa$ sensor consists of pixels sized $\unit{18.4}{\upmu\meter}\,\times\,\unit{18.4}{\upmu\meter}$, which are arranged in 1152 columns and 576 rows.
This adds up to a total of about six hundred thousand readout channels per sensor, covering an active area of about $\unit{10.6}{\milli\meter}\,\times\,\unit{21.1}{\milli\meter}$. 
The specifications of the $\Mimosa$ sensors quote a thickness of $\unit{50}{\upmu\meter}$. 
Measurements with a digital microscope reveal an average thickness over the six sensors used of $(54.5\,\pm\,3.6)\,\upmu\meter$. 
Free charge carriers produced in the underlying \unit{10}{\upmu\meter} high resistivity epitaxial layer\footnote{This is the case for DATURA and ACONITE.} 
 are collected via drift (diffusion) in depleted (undepleted) regions. 
The binary resolution of $\unit{5.3}{\upmu\meter}$ is improved by charge sharing, i.e.\ the collection of charge at adjacent pixels and subsequent calculation of the centre of gravity,
 as is shown in section~\ref{sec:trackres}.

The $\Mimosa$ sensors are read out with a rolling-shutter, taking 16\,cycles of an 80\,MHz clock per row, with all columns being read out in parallel. 
This allows for correlated double sampling and zero suppression on-chip with the digital circuitry placed outside the active pixel array. 
At this clock frequency, the $\Mimosa$ integration time equals $\unit{115.2}{\upmu\second}$, allowing for about 8680 frames to be read out per second. 
The expected maximum rate of detectable particles through the active area is estimated to be about $\unit{1}{\mega\hertz/\centi\meter^2}$ due to the limited on-chip buffer size. 
The detection threshold is programmable via so-called JTAG files. 
These provide configurations with different threshold levels in integer multiples $\noise$ of the RMS noise of the individual planes. 
A sensor threshold setting of $\noise = 6$ therefore corresponds to a collected charge in a single pixel of at least six times the noise. 
At this threshold, the average noise occupancy per pixel is measured to be about $6\cdot10^{-5}$ per readout frame at room temperature~\cite{ref:mimosa26}.

Every pixel sensor is mounted within an aluminium jig, three jigs are in turn mounted on each of the two aluminium arms. 
The jigs feature a beam window around the position of the sensor location, minimising the material budget. 
Lightproof Kapton foils of $\unit{25}{\upmu\meter}$ thickness protect the sensors on each side.
The overall material of the beam telescope thus amounts to $\unit{300}{\upmu\meter}$ of silicon and $\unit{300}{\upmu\meter}$ of Kapton. 
The two telescope arms, usually one up- and one downstream of the DUT, are movable along the direction of the beam in order to allow for variably sized DUTs to be fitted into the set-up. 
The minimal distance between two sensors is given by the jig thickness of 20\,mm, the maximal distance is restricted by the length of the aluminium arms to 150\,mm at equidistant spacing.
Furthermore, the jigs are cooled keeping the $\Mimosa$ sensors at a constant temperature of typically 18$\,\celsius$ for stable operation.
The entire beam telescope is placed on a rotatable frame easing adjustment of its orientation parallel to the beam. 
Additionally, this frame is mounted on a sturdy structure providing stability over time and wheels for easy transportation. 

\begin{figure}[tb]
	\center
	\ifdefined\notFOREPJ
	\includegraphics[width=.9\textwidth]{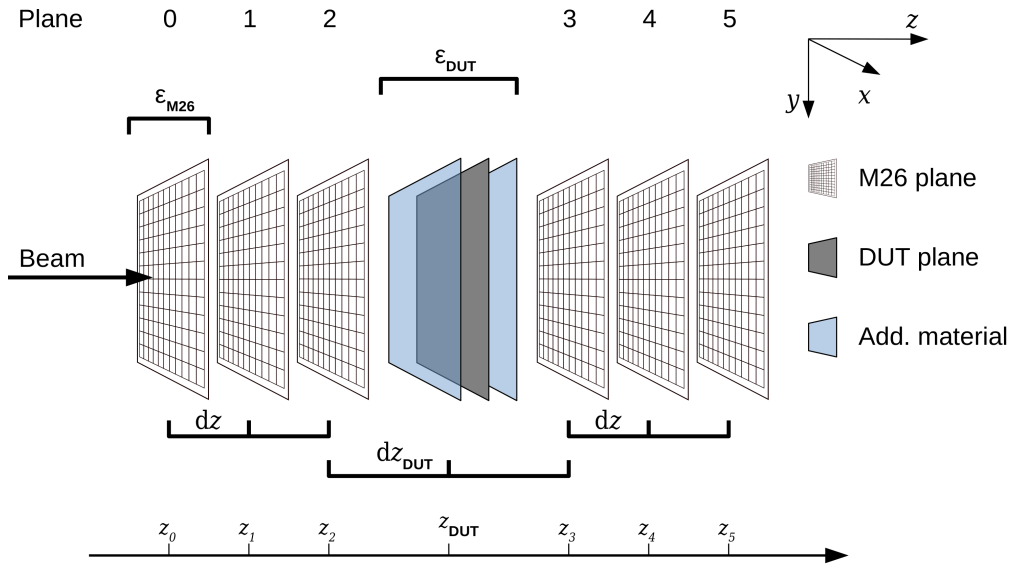}
	\else
	\includegraphics[width=.9\textwidth]{sketch_tscope6.eps}
	\fi
	\caption[Sketch of the $\DESY$-type telescope set-up]{Sketch of the $\DESY$-type telescope set-up and its important parameters.
	Planes 0 to 2 and planes 3 to 5 are referred to as upstream and downstream planes, respectively.}
	\label{fig:datura_sketch}
\end{figure}

Important parameters are compiled in figure~\ref{fig:datura_sketch}. 
The distance between $\Mimosa$ planes is denoted as $\dz$, the distance between the DUT and its nearest neighbouring $\Mimosa$ planes $\dzdut$. 
The $z$-positions of the pixel planes are denoted $z_0$ to $z_5$ and the DUT is placed at $z_{\textrm{DUT}}$. 
The expression $\varepsilon = \sum_i x_{i}/X_{0,i}$ defines the material budget of the scattering medium as the physical material thicknesses $x_i$ normalised to their radiation lengths,
 with values of $X_0 = 93.65\,\milli\meter$ for silicon, $X_0 = 3.042\cdot 10^{5}\,\milli\meter$ for dry air, and $X_0 = 285.6\,\milli\meter$ for Kapton, cf.~table~III.6 of reference \cite{ref:x0values}.
The material budget of a $\Mimosa$ plane including a protecting Kapton foil on each side is $\epsmimo = 7.5\cdot10^{-4}$ at an assumed sensor thickness of $\unit{54}{\upmu\meter}$. 
At a spacing of $\dz = \unit{20}{\milli\meter}$ ($\dz = \unit{150}{\milli\meter}$) the total material budget amounts to $4.8\cdot10^{-3}$ ($7.0\cdot10^{-3}$).
Likewise, $\epsdut$ represents the DUT's material budget which includes the DUT itself and other material such as PCBs and cooling boxes.

\subsection{Trigger and DAQ system}
\label{sec:tdaq}

Four Hamamatsu PMT assemblies with scintillators and lightguides, two in front and two in the back of the telescope, define the spatial acceptance for triggers. 
The crossed scintillators on either side of the beam telescope define a rectangular acceptance window of $10\,\milli\meter\,\times\,20\,\milli\meter$ matching the $\Mimosa$ sensor area. 
The TLU is based on a commercial Spartan\,3 board and features a coincidence unit with discriminator boards accepting up to four PMT input signals. 
Additionally, it is equipped with several custom-made add-on PCBs allowing for an easy integration of user DAQ systems. 
Providing a programmable logic, the TLU  takes a trigger decision based on its four input channels. 
As interface to the beam telescope DAQ and other user DAQs, the TLU provides RJ45 connectors with four LVDS pairs carrying the trigger clock, busy, reset, and trigger signals. 
The trigger clock and the busy line are inputs to the TLU, whereas the reset and the trigger line are signals produced by the TLU. 
The trigger signal is distributed to all DAQ systems on the trigger line. 
A busy signal is accepted by the TLU from each integrated DAQ individually vetoing subsequent triggers as long as the busy signal is high. 
More details are available in references~\cite{EUDET-2009-04,ref:TLUproc}.
In addition, a LEMO interface is available providing trigger and reset outputs and inputs for the busy signal.

Three different handshake modes handling the trigger/busy signals are available, one of them being the \textit{no-handshake mode},
 in which the TLU issues a fixed-length pulse on the trigger line with the busy line being disregarded. 
In \textit{simple handshake mode}, the assertion of a trigger is replied by the integrated DAQ systems by asserting a signal on the busy line. 
The TLU then de-asserts the trigger and waits for the busy line going low.
The \textit{trigger data handshake mode} uses the same scheme as in the \textit{simple handshake mode}, but additionally the current trigger data are transferred on the trigger line:
after the trigger has been de-asserted, the trigger number is clocked out. 
%

The zero-suppressed hit data generated by the $\Mimosa$ sensors are transmitted over a ribbon cable to the auxiliary boards,
 which establish the connection to the data concentrator board collecting the data from all six sensors and the trigger/busy lines from the TLU.
The signals from the concentrator board are acquired by a COTS\footnote{commercial off-the-shelf} DAQ system built around the National Instrument (NI) PXIe crate architecture. 
A custom-made firmware running on a Virtex 5 FPGA embeded in a FlexRIO board (PXIe 7962R) acquires and deserialises the 12 serial links -- two per sensor -- at 80\,Mb/s,
 detects triggers on the trigger line, and reads the trigger data. 
The resulting data stream of 960 Mb/s is read by the CPU (PXIe 8130) from the PXIe bus via a DMA channel and processed on-line by software. 
Data packets on which the TLU has triggered are selected, demultiplexed and actual frames built. 
The data are then available to the $\eudaq$ framework (cf.\ section~\ref{sec:eudaq}), and an event is written to disk in normal handshake mode, only if a trigger has been raised for a certain telescope readout frame. 
This co-development, i.e.\ sharing processing tasks between firmware and software, allows for high flexibility for system upgrades. 
The presented DAQ architecture is able to read six $\Mimosa$ sensors without any dead-time at up to 8680 frames/s.
More details are available in references~\cite{EUDET-2010-25,Claus}.

%% file: content/eudaq.tex
The modular cross-platform data acquisition framework $\eudaq$~\cite{ref:eudaqwebsite} has been designed and developed to serve as flexible and simple-to-use data taking software for the EUDET-type beam telescopes,
 allowing for easy integration of other devices. 
It consists of completely independent modules communicating via TCP/IP enabling a distributed data acquisition with modules running on separate machines. 
Currently, $\eudaq$ is designed for synchronous DAQ systems requiring one event per trigger per attached subdetector system before building the global event. 
Thus, the trigger rate is always limited by the slowest device.

The central interaction point for users with the framework is the Run Control and its graphical user interface (GUI). 
All other modules connect to the Run Control at startup and receive additional information from there during operation such as the commands for starting and stopping a DAQ run. 
The GUI provides all controls necessary to the user on shift. 
Another important user interface is the Log Collector, gathering logging information from all modules and displaying them in one unified logging window. 
Log files are stored along with the recorded data files for later reference.

The actual detector data are delivered to the framework by so-called Producers.
Producers are the links between the EUDAQ framework and the subdetector systems such as the beam telescope, the TLU, or user DAQ systems.
They interface with the EUDAQ library and provide a set of commands to be called by the Run Control. 
This simple interface scheme eases the integration of user DAQ systems into the software framework.
The data read out from the detectors by the individual Producers are sent to the so-called Data Collector. 
This Data Collector is responsible for the event building, i.e.\ the correlation of events from all subdetector systems to single global events comprising all data belonging to one trigger. 
Basic sanity checks such as testing the consistency of event numbers reported by the individual subdetectors are executed.

To ensure data quality during data acquisition, the Online Monitor tool is available. 
It connects to the Data Collector requesting a fixed fraction of the recorded events (e.g.\ one out of a hundred) to fully decode all subdetector data
 and build basic plots such as hit maps or correlation plots.
Thereby, data quality monitoring verifying that the different devices are synchronized in time and all within the geometrical trigger acceptance, is possible during data taking.

The data decoding is performed using Data Converter plug-ins for every detector type attached to the Run Control. 
The plug-in to be called for a specific subevent is deduced from the event type transmitted by the Producer and written to the data stream by the Data Collector. 
Each Data Converter plug-in can implement several data format end points allowing, amongst others, for the conversion to the internal EUDAQ format for the Online Monitor, to simple ROOT trees, or to LCIO
 which is used by the offline reconstruction software described in the following section.

Configuration of the data acquisition framework is performed via global configuration files. 
The information for every individual module is parsed and distributed by the Run Control. 
The configuration file is a plain text file divided into sections for the individual modules, which contain a list of parameter-value pairs defined by the module.
The full content of the configuration file including commented lines is stored in the so-called Begin-Of-Run Event (BORE) of every run and is thus available later for offline analysis and reference. 
This greatly simplifies book keeping of detector parameters during test beam shifts since all settings are stored automatically.

%% file: content/offline.tex
For offline analysis and reconstruction of telescope test beam data the EUTelescope software package~\cite{ref:eudetmemo_2010_12,ref:eutelwebsite}
 is available and features a close integration of the EUDAQ software framework described in section~\ref{sec:eudaq}.
EUTelescope is based on the ILCSoft framework~\cite{ref:eudetmemo_2009_12} which provides the basic building blocks for offline analysis such as a generic data model (Linear Collider I/O, LCIO),
a geometry description language (GEAR) and the central event processor (Marlin) \cite{ref:eudetreport_2007_11}.

Marlin allows for a modular composition of analysis chains for various applications. Every task is implemented as an independent processor which is called by Marlin for every event. 
Each processor exposes a set of parameters to the user which can be configured and loaded at runtime via so-called steering files in XML format.
This way the Marlin/Processor architecture gives maximum flexibility to the user.

\begin{figure}[tbp]
  \center
  \ifdefined\notFOREPJ
  \includegraphics[width=.9\textwidth]{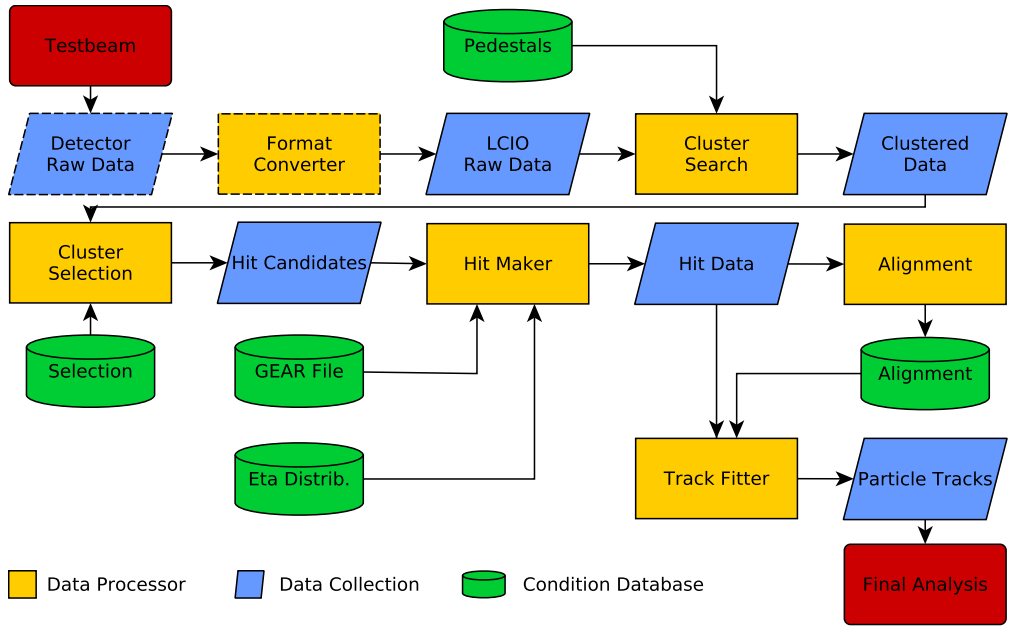}
  \else
  \includegraphics[width=.9\textwidth]{eutel-strategy.eps}
  \fi
  \caption[The EUTelescope data analysis strategy]{Schematic of the overall telescope data reconstruction and analysis strategy of the EUTelescope framework.
EUTelescope provides processors for all steps, except for the conversion of the DUT raw data, marked with a dashed outline.}
  \label{fig:offline:strategy}
\end{figure}

EUTelescope provides several processors for Marlin, implementing algorithms necessary for a full track reconstruction and data analysis of test beam experiments. 
Figure~\ref{fig:offline:strategy} shows the analysis strategy of the framework starting from the recorded detector response to the final reconstructed particle tracks. 
At low-energy beam lines such as the DESY-II test beam facility, multiple scattering is an important contribution to the overall track resolution uncertainty,
 especially in measurements with non-negligible DUT material budget, cf.\ section~\ref{sec:resmultiple}.
Therefore, EUTelescope provides processors implementing advanced algorithms for tracking based on the concepts of a Deterministic Annealing Filter (DAF)~\cite{ref:daffitter}
 or General Broken Lines (GBL)~\cite{Blobel20111760,Kleinwort-2012}, which account for scattering in all material present in the beam. 
For high-energy beam lines a simple straight line fit provides sufficient precision and a maximum of computational performance by employing a $\chi^{2}$-minimisation method~\cite{ref:eudetmemo_2007_01,ref:lutzpaper} to calculate the track parameters.
In addition, precise offline detector alignment can be performed by minimising track residuals using the EUTelescope alignment processor which utilises the Millepede-II algorithm~\cite{Blobel-2006}.

EUTelescope comes with its own job submission framework Jobsub that allows for submission of analysis jobs on local machines or on larger batch computing clusters such as the NAF\footnote{National Analysis Facility. This is a German UNIX cluster.}
 or LXPLUS\footnote{LXPLUS is a CERN UNIX cluster.} for bulk reconstruction.
Using its flexible configuration file concept and the global run database storing user defined variables,
 Jobsub eases the implementation of per-run variables for reconstruction such as beam energy or detector alignment.

%% file: content/track_resolution.tex
The data presented were taken in $2015$ using $\Datura$, an $\eudet$-type beam telescope, which was operated with $\eudaq$, cf.~sections~3 and~4. 
Results shown in this section have been obtained using the $\EUTelescope$ software described in section~\ref{sec:offline}. 
The set-up does not include any additional DUT, only six $\Mimosa$ telescope planes. 
Track fits are performed using the GBL formalism and track residuals, i.e.\ the distance between the track fit and the measured hit,
 are calculated for each telescope plane. 
The intrinsic resolution of the sensor planes is derived from data and used as input for track resolution predictions.

\subsection{Data analysis flow and pattern recognition}
\label{sec:datura-nodut}

After conversion from raw $\Mimosa$ data to the LCIO data format, a hot pixel search is performed marking pixels with firing frequencies above a certain threshold.
Clusters are formed from fired, adjacent pixels and translated from two-dimensional entities on the individual telescope planes into hits in a global three-dimensional reference frame.
Clusters containing at least one hot pixel are removed from the analysis. 
Triplets are built from hits in the three upstream and three downstream planes separately: 
First, doublets are defined by straight lines from all hits in plane\,0 to all hits in plane\,2. 
A valid triplet is found, if a matching hit in plane\,1 is present, non-valid triplets are discarded.  
Triplet isolation is ensured by rejecting all triplets whose extrapolation at a chosen $z$-coordinate approaches other triplet extrapolations closer than $\unit{300}{\upmu\meter}$,
 the $z$-coordinate being the centre between plane\,2 and plane\,3. 
The procedure is repeated for the downstream planes. 
Matching triplets from the up- and downstream planes are identified by isolated triplets that intersect within a radius of $\unit{100}{\upmu\meter}$, again at the centre between plane\,2 and plane\,3. 
In turn, GBL tracks are formed by the six hits belonging to the matched triplets. 
For offline alignment, the GBL tracks are passed to \mbox{Millepede-II} in order to determine shift and rotation alignment constants for every telescope plane.
Alignment constants of the sensitive coordinates are determined to submicron precision and are hence much smaller than the intrinsic resolution of the detectors used. 
After applying the alignment corrections, the above track finding is repeated, and GBL tracks are calculated based on the corrected hits. 
These tracks can be formed in a \textrm{biased} or in an \textrm{unbiased} way, where biased tracks include the measured hit information of every plane in contrast to unbiased tracks,
 that omit the hit information of the plane under investigation. 
Finally, a fiducial cut discards tracks at the border of the sensors and a cut on the $\chi^2$-probability for the given number of degrees of freedom rejects track with outliers from the further analysis. 

\subsection{GBL track fits, multiple scattering and residuals}
\label{sec:resmultiple}

The figure of merit for a beam telescope is its track resolution\footnote{In the literature, this is also referred to as \textit{pointing resolution}.} $\sigmatb$ ($\sigmatu$).
It defines the precision with which a particle trajectory can be determined for a biased (unbiased) track. 
This quantity is a function of the track path, i.e.\ it is not constant along the trajectory. 
It depends on the intrinsic resolution $\sigmai$ of the sensors, that have been used to measure the hits belonging to the track, the number of measurements and their positions
 as well as the amount of multiple scattering of the beam particles.
 
Various methods exist to perform the calculation of the track resolution. 
In the formalism of GBL, an initial simple trajectory (straight line, helix) is corrected for multiple scattering. 
A single, global $\chi^2$-fit of a track model is performed involving all measured hits including their measurement uncertainties (intrinsic resolution)
 and scattering angle uncertainties along the trajectory. 
For tracks not bent by a magnetic field, this results in trajectories formed of straight lines between defined scatterers allowing for kinks only at the scatterers themselves. 
The track resolution stems from the covariance matrix, which is calculated for each track parameter~\cite{Blobel20111760,Kleinwort-2012}. 
The method of GBL makes various assumptions and simplifications:

\begin{itemize}
 \item Scatterers, the origin of kinks in the trajectory, are thin, i.e.\ no spatial offsets within a scatterer are taken into account. 
 \item A thick scatterer, like the air between two planes, is modelled by two thin scatterers. 
 \item The offsets of the local track parameters are small and are propagated linearly. 
 \item All random variables follow normal distributions.
 \item Bremsstrahlung effects are not taken into account in the track model. 
\end{itemize}
 
%
%
%
%
%
%
Multiple Coulomb scattering causes angular deflection of charged particles traversing any medium.
The angular scattering distribution is centred around zero
 and the standard deviation depends on the particle energy, particle type and the radiation length of the matter traversed~\cite{ref:scatteringhighland}.
An approximation for high energy protons yields a closed form for the variance of the angular distribution. 
In the transverse plane, the variance for a single scatterer $\varepsilon$ reads~\cite{ref:PDG-2014}

\begin{equation}
\label{eq:multiplescattering_a}
\Theta_{0}^{2} = \left( \frac{13.6\,\mega\electronvolt}{\beta c p} \cdot z \right)^2
\cdot \varepsilon
\cdot \left( 1 + 0.038 \cdot \ln{\left( \varepsilon \right) } \right)^2 \,,
\end{equation}

\noindent with the velocity $\beta c$, the momentum $p$ and the charge number $z$ of the traversing particle. 
An accuracy of 11\,\% for all atomic numbers $Z$ is reported for the standard deviation of the angular distribution. 
For a composition of scatterers $\varepsilon_i$, cf.\ section~\ref{sec:sensors}, the variance including all scatterers (i.e.\ after the last scatterer)
 involves the sum over the material budgets and hence $\varepsilon$ in equation~(\ref{eq:multiplescattering_a}) is to be replaced by $\varepsilon = \sum \varepsilon_i$. 
For track fitting (with a Kalman filter or GBL) the variance from all scatterers along the track has to be distributed over the individual scatterers. 
Using the material fraction as relative weight results in a formula similar to equation~(\ref{eq:multiplescattering_a})
 with the total material budget used in the logarithmic correction

\begin{equation}
\label{eq:multiplescattering}
\Theta_{0,i}^{2} \equiv \dfrac{\varepsilon_i}{\varepsilon} \cdot \Theta_{0}^{2} = \left( \frac{13.6\,\mega\electronvolt}{\beta c p} \cdot z \right)^2
\cdot \varepsilon_i
\cdot \left( 1 + 0.038 \cdot \ln{\left( \varepsilon \right) } \right)^2 \,.
\end{equation}

The amount of multiple scattering is calculated per sensor considering the measured sensor thickness and a $25\,\micro\meter$ thick Kapton foil on either side of each sensor.
The contribution of air between adjacent planes is accounted for as additional scatterers. 
Equation~(\ref{eq:multiplescattering}) demonstrates the advantage of using thin sensors, since the standard deviation of the angular distribution increases with the material budget.
This is especially important at low-energy beams, such as the DESY-II test beam, as the standard deviation increases towards lower beam energies.

The biased (unbiased) residual defined as a quantity per track is the distance between the measured hit and the biased (unbiased) track fit. 
The standard deviation of the residual distribution is hereafter denoted as the \textit{residual width}. 
With a known or estimated biased track resolution, i.e.\ all $\sigmai(z_i)$ are known or estimates exist,
 the \textrm{biased} residual width for a set of tracks can be expressed for all positions $z = z_i$ as

\begin{equation}
 \label{eq:telescoperesolutionequation1}
 \rbiased^2(z) = \sigmai^2(z) - \sigmatb^2(z).
\end{equation}

\noindent
The biased residual width at a considered plane $i$ decreases with increasing plane distance $\dz$ and towards lower beam energies. 
These effects stem from the term $\sigmatb$: 
qualitatively, the biased track resolution increases, i.e.\ worsens, with larger lever arms $\dz$, as an uncertainty in the deflection angle is boosted with larger distances. 
It converges towards the intrinsic resolution in the limit of large $\Theta_0$. 
Hence, the difference between $\sigmai$ and $\sigmatb$ decreases. 
This is also true when comparing residuals from the inner planes to those from the outer ones:
The worsening track resolution towards the outer planes decreases the difference with respect to $\sigmai$ and hence the residual width. 

If the hit measurement of a telescope plane at position $z_i$ is not included in the fit,
 the \textrm{unbiased} residual width at these positions reads~\cite{ref:eudetreport200902}
 
\begin{equation}
\runbiased^2(z) = \sigmai^2(z) + \sigmatu^2(z).
\label{eq:telescoperesolutionequation2} 
\end{equation}

\noindent
Contrary to a biased residual, the unbiased residual at a considered plane $i$ increases with increasing plane distance $\dz$ and towards lower beam energies. 
The unbiased track resolution also increases with larger lever arms $\dz$, but is not limited by the intrinsic resolution. 
Hence, the sum of $\sigmai$ and $\sigmatu$ increases. 
Both the biased and the unbiased residual distributions are a function of the track resolution and therefore feature the same dependencies.

The biased pull of a track is defined as the ratio of the biased residual over the predicted biased residual width

\begin{equation}
 \textrm{pull}_{\textrm{b}} \equiv \pb = \frac{\rbiased}{\sqrt{\sigmai^2 - \sigmatb^2}}.
 \label{eq:pull}
\end{equation}

\noindent
Assuming normally distributed residuals, the biased pull distributions are normally distributed and centred at zero with a standard deviation of one, $N(0,1)$,
 if the material and the scattering therein is known and correctly described. 
A standard deviation differing from one originates from an under- or overestimated residual prediction $\sqrt{\sigmai^2 - \sigmatb^2}$,
 which in turn stems from an inaccurate estimate of either the intrinsic resolution or the track resolution. 
The latter is a consequence of inaccurate inputs for the standard deviation $\Theta_0$ of the assumed angular distribution or again for the intrinsic resolution.

\subsection{Methodology and systematics}
\label{sec:meth}

\begin{table}[tbp]
\caption[Systematic uncertainties]{Systematic uncertainties of the intrinsic resolution listed for biased and unbiased track fits and two geometries at 6\,GeV and 2\,GeV.}
 \begin{center}
  \begin{tabular}{r|r|r|c|c|c|c||c}
  \multicolumn{3}{c|}{} & \multicolumn{5}{c}{$\sigma_{\sigmai}$ in \%}\\
  \multicolumn{3}{c|}{} & \multicolumn{3}{c|}{per plane} & all planes& $\sqrt{\sum (x_i)^2}$\\
  \multicolumn{3}{c|}{} &$E$         & $\Theta_0$      & fit range    & rms($\pb$) &  \\ 
  \multicolumn{3}{c|}{} & $\pm5\,\%$ &  $\pm 3\,\%$  & $\pm 1\,$std.&               &  \\ \hline
  6\,GeV & 20 mm   &  biased  &  ${}^{-0.34}_{+0.21}$ & ${}^{+0.08}_{-0.28}$ & ${}^{+1.76}_{-1.27}$ & 1.57 &  2.6  \\
         & ×       & unbiased &  ${}^{-0.43}_{+0.71}$ & ${}^{+0.44}_{-0.25}$ & ${}^{-0.93}_{-1.00}$ & 1.23 &  1.8  \\
	 & 150 mm  &  biased  &  ${}^{-3.5}_{+2.9}$   & ${}^{+1.95}_{-2.60}$ & ${}^{+6.4}_{-5.4}$   & 1.51 &  7.9  \\
	 & ×       & unbiased &  ${}^{-4.80}_{+5.43}$ & ${}^{+2.97}_{-4.13}$ & ${}^{-5.29}_{+3.11}$ & 0.75 &  8.7  \\ \hline
  2\,GeV & 20 mm   &  biased  &  ${}^{-1.56}_{+1.13}$ & ${}^{+0.65}_{-1.22}$ & ${}^{+0.23}_{+0.33}$ &  3.1 &  3.7  \\
         & ×       & unbiased &  ${}^{-1.67}_{+1.21}$ & ${}^{+0.92}_{-1.10}$ & ${}^{-2.15}_{+1.35}$ & 1.94 &  3.1  \\
	 & 150 mm  &  biased  &  ${}^{-10.5}_{+15.7}$ & ${}^{+10.2}_{-6.59}$ & ${}^{+8.0}_{+0.82}$  & 0.82 &  20.3 \\
	 & ×       & unbiased &  ${}^{-17.5}_{+24.9}$ & ${}^{+14.9}_{-15.2}$ & ${}^{-23.9}_{+25.1}$ & 1.03 &  38.5  \\ 
  \end{tabular}
  \label{tab:uncerts}
 \end{center}
\end{table}

As the intrinsic resolution is not known a priori, an initial estimate, $\sigmahat$, is used as an input to the GBL track fitting. 
The intrinsic resolution $\sigmai$ is assumed to be constant over the active area of the sensor and is averaged over all cluster sizes. 
The standard deviation of the pull distributions is iteratively used to update the estimate $\sigmahat$. 
The same estimate is used for all six planes. 
The calculation is repeated for five beam energies between 6\,GeV and 2\,GeV, ten different sensor thresholds
 and two different geometries: the narrow geometry with $\dz = \dzdut = 20\,\milli\meter$ and the wide geometry with $\dz = \dzdut = 150\,\milli\meter$.

This analysis compares results between biased and unbiased track fits. 
In principal, both choices lead to compatible results if the material in the beam and the deflection due to multiple scattering therein is correctly described. 
Systematic uncertainties affect the track resolution, the predicted residual width, and finally the derived intrinsic resolution. 
The systematic uncertainties of the intrinsic resolution are calculated using the same iterative method used for the central value and
 are estimated based on uncertain beam energy, uncertain standard deviation of the angular distribution and variations in the fit range. 
The beam energy is varied by 5\,\%. 
The original systematic uncertainty of the standard deviation $\Theta_{0}$ of $11\,\%$ is experimentally constrained to 3\,\%, as is shown in section~\ref{sec:measurements}. 
All systematic uncertainty contributions for the two set-ups are listed in table~\ref{tab:uncerts}. 
The impact of systematic uncertainties on the result is similar between the biased and the unbiased methodology. 
In addition to the above mentioned contributions per plane,
 there is a systematic uncertainty of the method itself, as it predicts slightly different intrinsic resolutions for each plane, which changes between geometries,
 i.e.\ it is not dominantly caused by an actual difference in intrinsic resolution. 
The impact of the variance between the width of the pull distributions for different sensors on the intrinsic resolution is therefore included in the table. 
The most accurate measurement of the intrinsic resolution can be performed at minimal multiple scattering, i.e.\ at 6\,GeV and $\dz = 20\,\milli\meter$,
 featuring a total systematic uncertainty of $2.6\,\%$ and $1.8\,\%$ for biased and unbiased track fits, respectively. 
Measurements at 2\,GeV and $\dz = 150\,\milli\meter$ yield an uncertainty of 20\,\% to 40\,\%. 

The assumptions made in the GBL formalism might be less valid for measurements towards lower energies and larger plane spacings, i.e.\ measurements with a large amount of multiple scattering. 
Therefore, the predicted track resolution is systematically affected, and hence the pull distributions. 
Consequently, the estimator $\sigmahat$ does not converge towards the true $\sigmai$. 
Therefore, a global energy dependent tuning parameter $\kappa$ is introduced,
 that allows for an adapted amount of scattering used as input to GBL,

\begin{equation}
 \Theta_{\textrm{corr}} = \kappa(E) \cdot \Theta_0.
 \label{eq:thetacorr}
\end{equation}

\noindent
This parameter is assumed to be constant as a function of the sensor threshold and chosen geometry. 
Since the intrinsic resolution is a priori independent of the plane spacing $\dz$,
 a comparison at two precisely measured spacings allows for a calibration of the amount of scattering used as input to GBL at a given beam energy. 
The value $\kappa$ is determined iteratively under the constraint that the resulting intrinsic resolution for the two geometries agree within 10\,nm.
The uncertainty of $\kappa$ is defined as the range for which the intrinsic resolution of the wide geometry equals the central value plus/minus one standard deviation of the narrow geometry
 excluding the uncertainty on $\Theta_0$. 
If $\kappa$ differs significantly from one and/or depends on the beam energy is to be determined with data in section~\ref{sec:measurements}. 
Also, a possible difference between the biased and the unbiased fit method is investigated.

\subsection{Performance measurements with DATURA}
\label{sec:measurements}

\begin{figure}[tbp]
  \centering
  \ifdefined\notFOREPJ
  \includegraphics[width=0.49\textwidth]{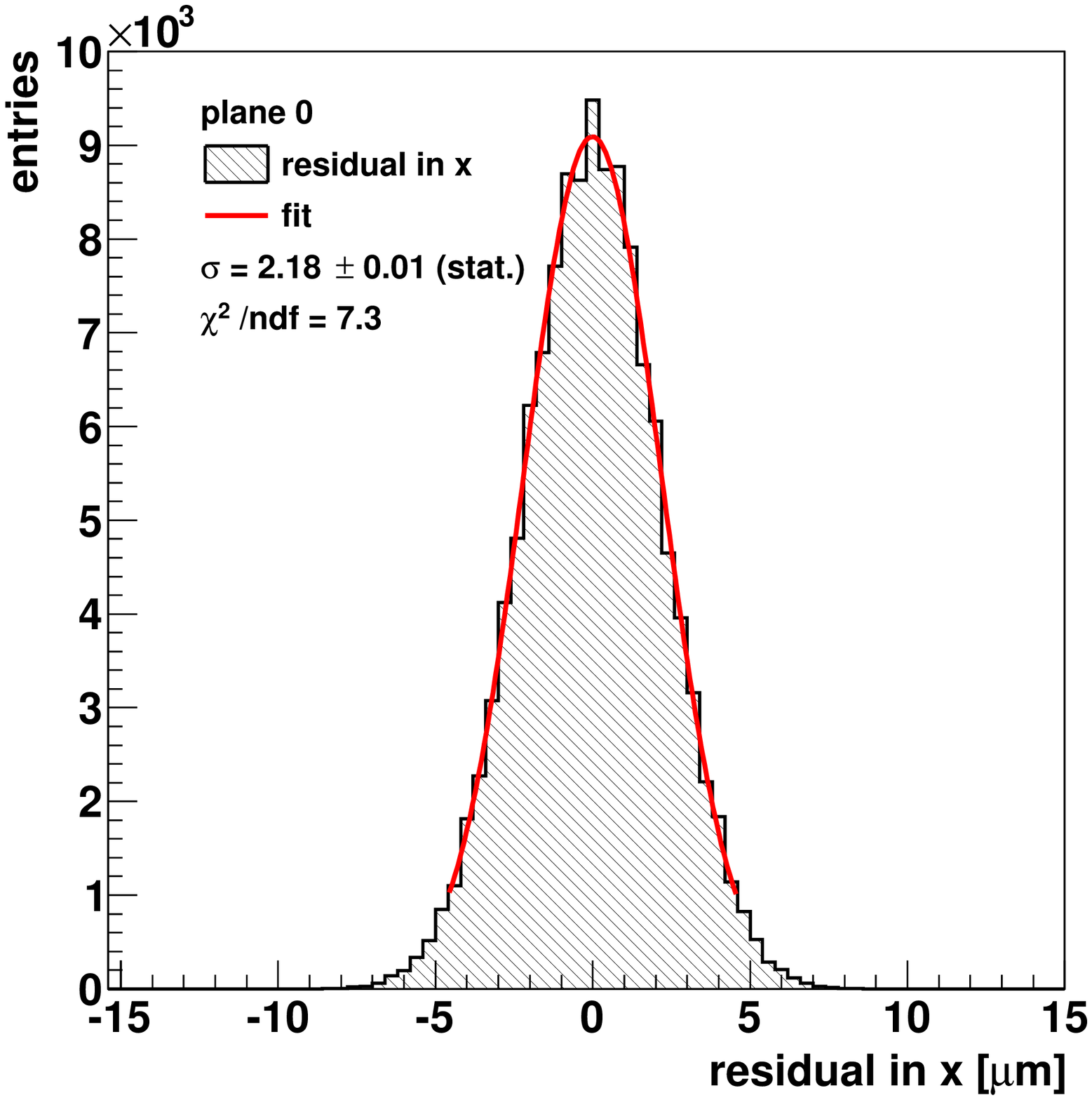} \put(-30,158){(A)}
  \includegraphics[width=0.49\textwidth]{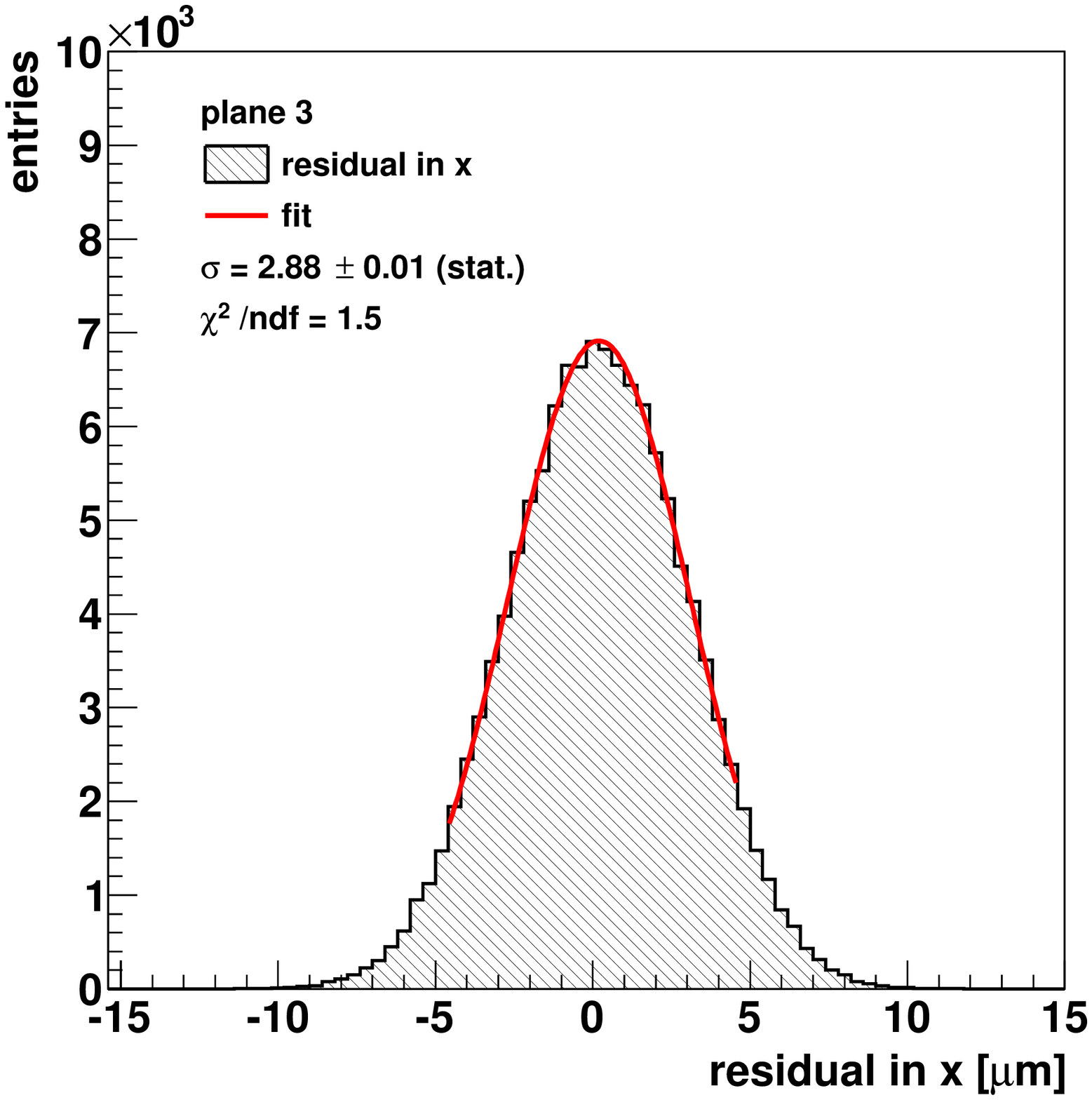} \put(-30,158){(B)}\\
  \else
  \includegraphics[width=0.49\textwidth]{0x.eps} \put(-30,158){(A)}
  \includegraphics[width=0.49\textwidth]{3x.eps} \put(-30,158){(B)}\\
  \fi
  \caption[Residual examples to determine the $\Datura$ telescope's resolution]{
  Biased residual distributions measured with the $\Datura$ telescope at 6\,GeV with a plane spacing of $\dz = 20\,\milli\meter$. 
  The measured residuals in the $x$-direction for plane $0$ (A) and plane $3$ (B) are shown.}
  \label{fig:residualexample1}
\end{figure}

\noindent
Measurements of particle trajectories were performed to verify the performance of the $\Datura$ telescope at various beam energies, sensor thresholds, and sensor spacings. 
The residual and pull distributions have been obtained for every sensor plane as described in sections~\ref{sec:datura-nodut}, \ref{sec:resmultiple}, and~\ref{sec:meth}
 for both methodologies, biased and unbiased track fits~\cite{jansen_2016_49065}. 
%



Figure~\ref{fig:residualexample1} shows the biased residual distribution in $x$ for plane\,0 and plane\,3 for a telescope sensor spacing of $\dz = 20\,\milli\meter$,
 a beam energy of 6\,GeV and a sensor threshold setting of $\noise = 6$. 
According results for the $y$-direction, further planes, the wide geometry and the unbiased methodology are omitted. 
The distributions are fitted with a Gaussian, from which the residual width $\rbiased$ is determined. 
For plane\,3, the biased residual width in the $x$-direction is 

\begin{equation}
\left( 2.88\,\pm\, 0.01\,\textrm{(stat.)}\,\pm\, 0.08\, \textrm{(sys.)}\right)\,\micro\meter.
\end{equation}

\noindent
As expected, cf.\ section~\ref{sec:resmultiple}, the biased residual width for the outer plane\,0 is smaller than the width obtained from plane\,3.
Since the systematic uncertainty dominates the total uncertainty, the statistical contribution is neglected for the further analysis. 
It should be noted that the residuals feature non-Gaussian tails, as is expected from the underlying physics of the scattering mechanism~\cite{ref:PDG-2014}. 
The measured residual width used in this work is defined as the standard deviation of a Gaussian fit on the centre $95.5\,\%$ of the residual distribution.

%

\begin{figure}[btp]
  \centering
  \ifdefined\notFOREPJ
  \includegraphics[width=0.49\textwidth]{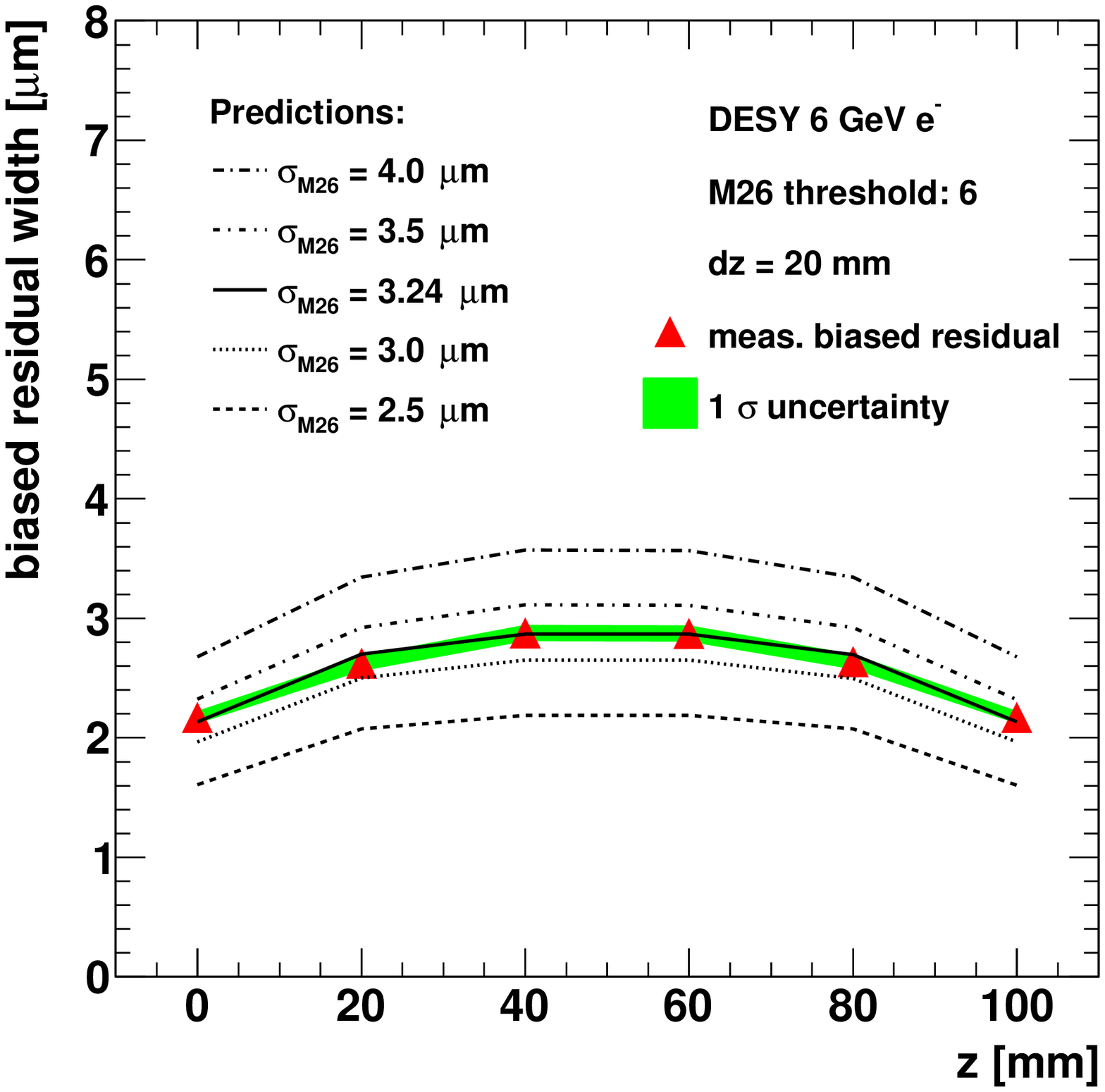}  \put(-30,158){(A)} 
  \includegraphics[width=0.49\textwidth]{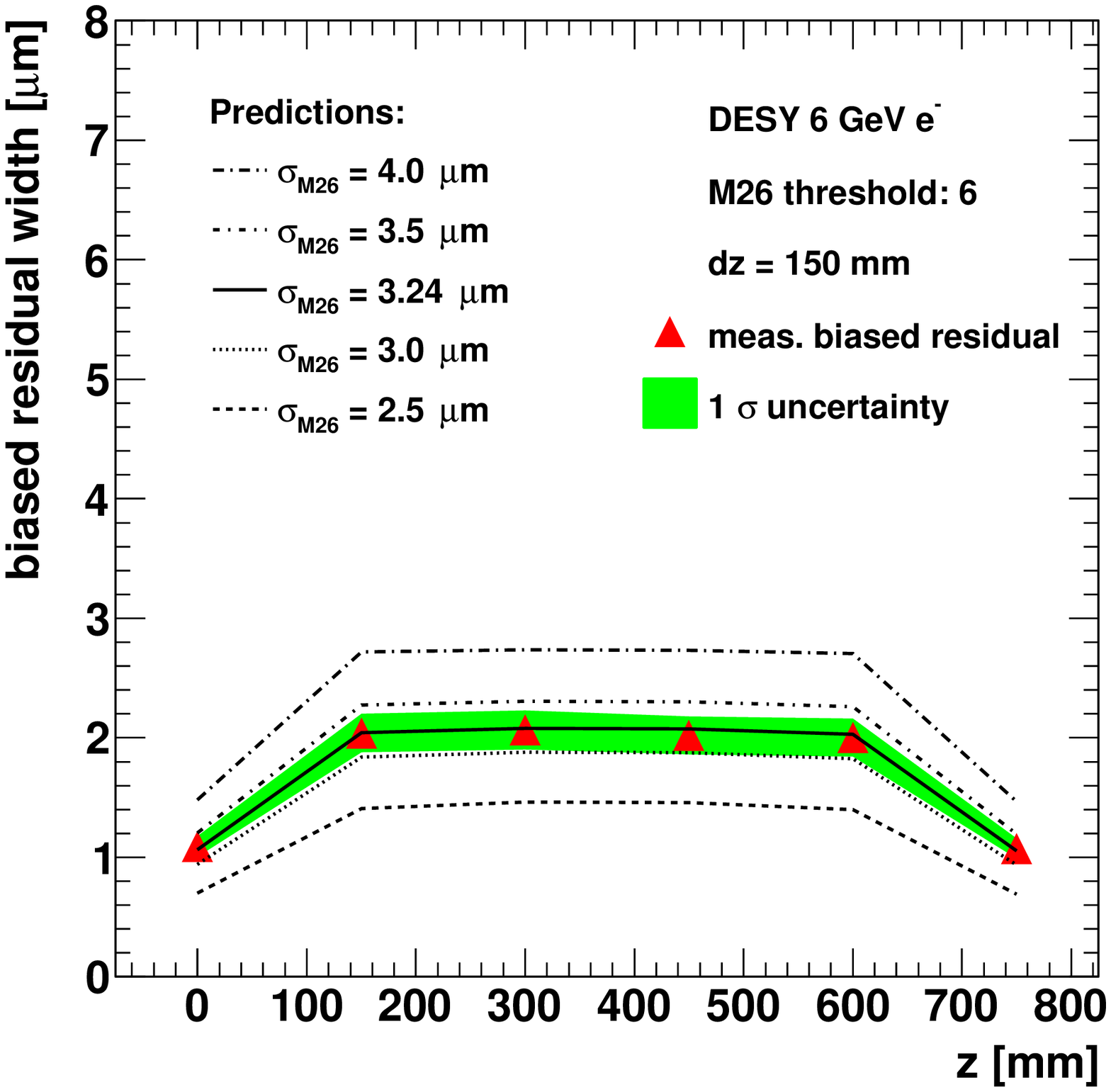} \put(-30,158){(B)} 
  \else
  \includegraphics[width=0.49\textwidth]{res_vs_z_20.eps}  \put(-30,158){(A)} 
  \includegraphics[width=0.49\textwidth]{res_vs_z_150.eps} \put(-30,158){(B)} 
  \fi
  \caption[The measured residual widths of each telescope plane.]{
  The measured residual widths of each telescope plane are shown in the $x$-direction for a plane spacing of $\dz = 20\,\milli\meter$ (A) and $\dz = 150\,\milli\meter$ (B).
  The black line shows the predicted residual width at $\sigmam = 3.24\,\upmu\meter$, the green band the measurement standard deviation.
  }
  \label{fig:smiley}
\end{figure}

The six measured residual widths in the $x$-direction at a beam energy of 6\,GeV and a sensor threshold $\noise = 6$ for both the narrow plane spacing of $\dz = 20\,\milli\meter$
 and the wide spacing of $\dz = 150\,\milli\meter$ are shown in figure~\ref{fig:smiley} (A) and (B), respectively. 
The two distributions show a global maximum in the centre with decreasing residual widths towards both ends. 
At a considered plane $i$ the residuals are smaller for the wide configuration in comparison to the narrow one, cf.~section~\ref{sec:resmultiple}.

As described in section~\ref{sec:meth}, the tuned amount of multiple scattering used as input to GBL
 is determined by varying $\kappa$ iteratively until the resulting resolution for the two considered geometries agree. 
This procedure results in an estimate of both $\kappa$ and the true intrinsic resolution. 
At 6\,GeV and 2\,GeV, a value of $\kappa(6\,\textrm{GeV}) = 0.76 \pm 0.04$ and $\kappa(2\,\textrm{GeV}) = 0.75 \pm 0.02$ is found, the latter corresponding to a relative uncertainty of $3\,\%$,
 which is used as the uncertainty on $\Theta_0$. 
Additionally, the average intrinsic resolution of the $\Mimosa$ sensors at 6\,GeV and threshold $\noise = 6$ results to
\begin{equation}
 \sigma_{\textrm{M26}} = \left( 3.24\,\pm\, 0.09 \right)\,\micro\meter.  
\end{equation}

\noindent
This average is taken over all twelve measurements of the telescope (two dimensions per plane). 
The result is consistent with earlier analyses, which yielded intrinsic resolutions of $\sigmai \approx\,3.4\,\micro\meter$ \cite{ref:thomas} and $\sigmai \approx\,3.5\,\micro\meter$~\cite{ref:mimosa26}.

\begin{figure}[tb!]
  \centering
  \ifdefined\notFOREPJ
  \includegraphics[width=0.49\textwidth]{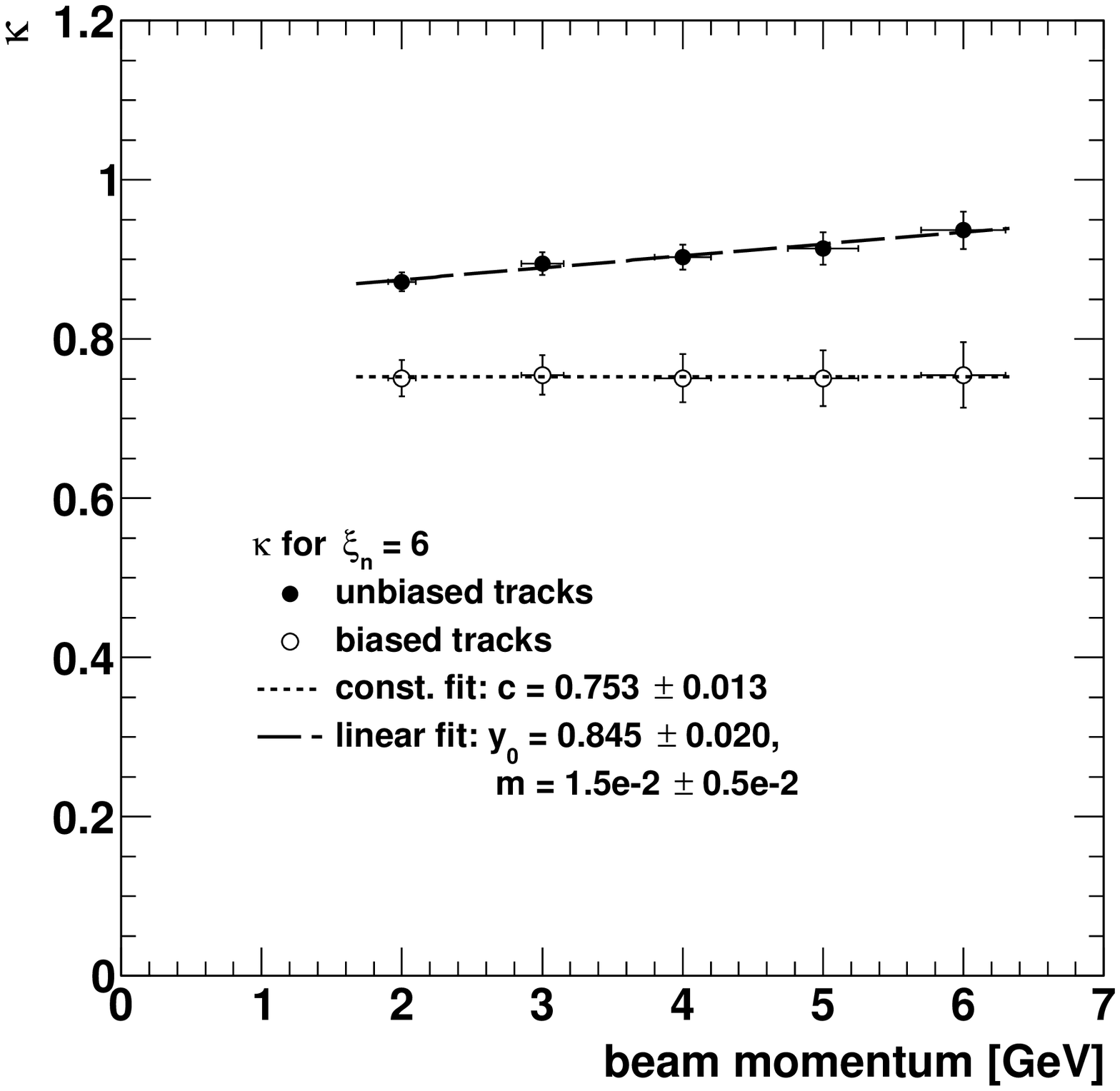}\put( -35,32){(A)}
  \includegraphics[width=0.49\textwidth]{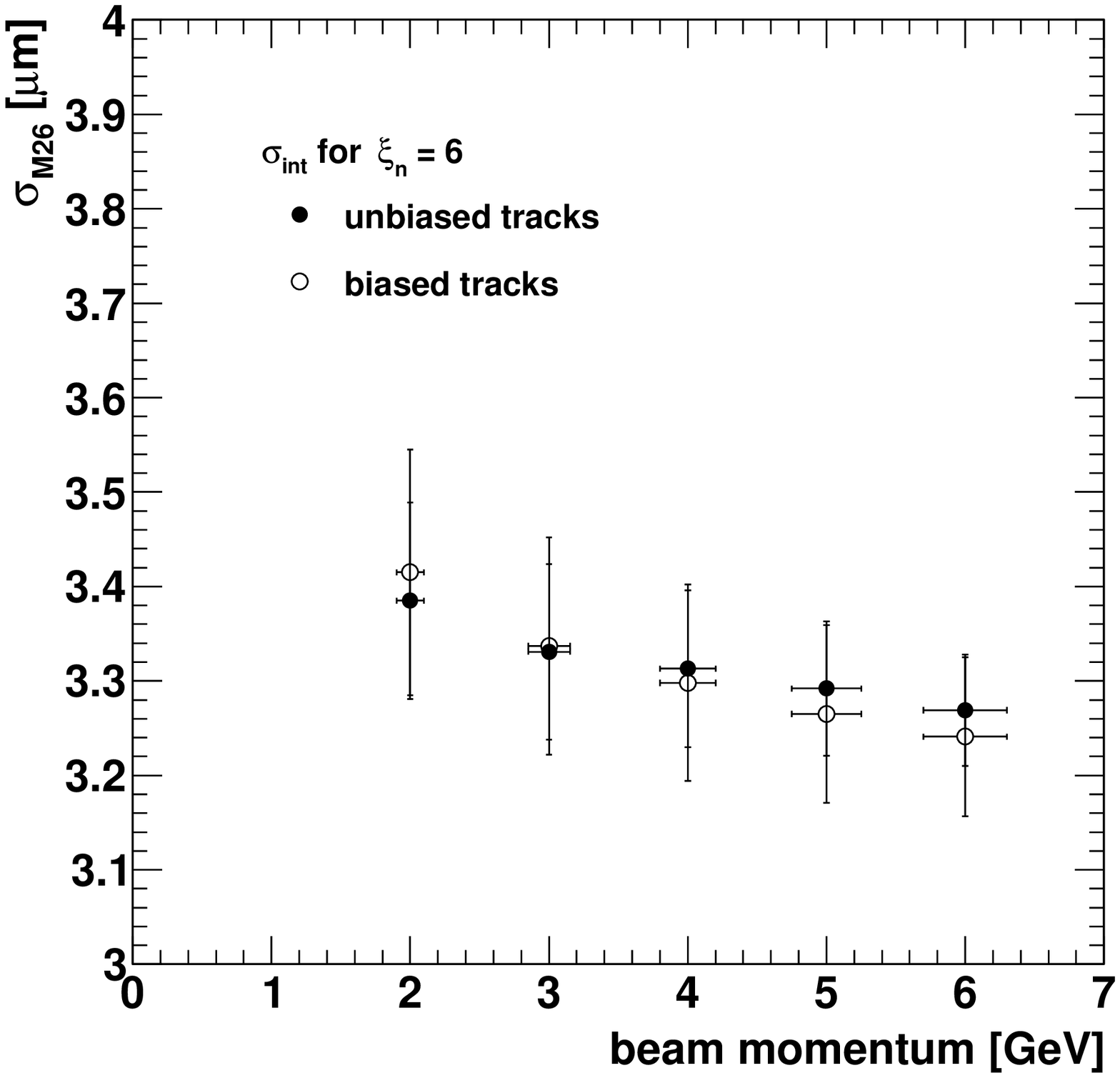}\put( -35,32){(B)}
  \else
  \includegraphics[width=0.49\textwidth]{kappa.eps}\put( -35,32){(A)}
  \includegraphics[width=0.49\textwidth]{sigmaint_ub.eps}\put( -35,32){(B)}
  \fi
  \caption[HL Factor]{
  (A) The global tuning factor $\kappa$ is plotted against beam energy. (B) The derived intrinsic resolution for biased and unbiased track fits.
  }
  \label{fig:HL_factor}
\end{figure}

The above method is repeated for different beam energies at a fixed threshold $\noise = 6$. 
This allows for the determination of $\kappa$ as a function of the beam energy. 
The correlation is shown in figure~\ref{fig:HL_factor}~(A) for both biased and unbiased track fits. 
For biased track fits, $\kappa$ is found to be constant at a value of $\kappa = 0.753 \pm 0.013$ in the energy range covered,
 which translates to a tuned input of about $-25\,\%$ with respect to equation~(\ref{eq:multiplescattering}). 
For unbiased track fits, $\kappa$ slightly increases from 0.87 at 2\,GeV to 0.94 at 6\,GeV with a precision of a single measurement of about $2\,\%$.
Figure~\ref{fig:HL_factor}~(B) shows the measured intrinsic resolution for both methodologies as a function of the beam energy. 
At each measured energy, they agree within less than one standard deviation over the entire energy range. 
The averaged absolute values vary slightly with energy, from $3.26\,\micro\meter$ at 6\,GeV to $3.40\,\micro\meter$ at 2\,GeV. 

Furthermore, the method has been repeated for various sensor thresholds.
The threshold applied to each telescope sensor is a critical parameter for the telescope performance.
A higher threshold reduces the amount of fired pixels which results in smaller and fewer clusters found on average on a plane and thus limiting the number of reconstructible tracks.
This reduces the telescope's triplet efficiency, which is defined as the ratio of isolated, upstream triplets with a matching hit on plane\,3 within an acceptance range $d$
 around the triplet extrapolation to the overall number of isolated, upstream triplets
 
\begin{equation}
 \epsilon_{\textrm{trip}} = \frac{\#\textrm{iso., upstream triplets} \cap \textrm{matching hit on plane 3 }}{\#\textrm{iso., upstream triplets}}.
\end{equation}

\begin{figure}[t]
  \centering
  \ifdefined\notFOREPJ
  \includegraphics[width=0.49\textwidth]{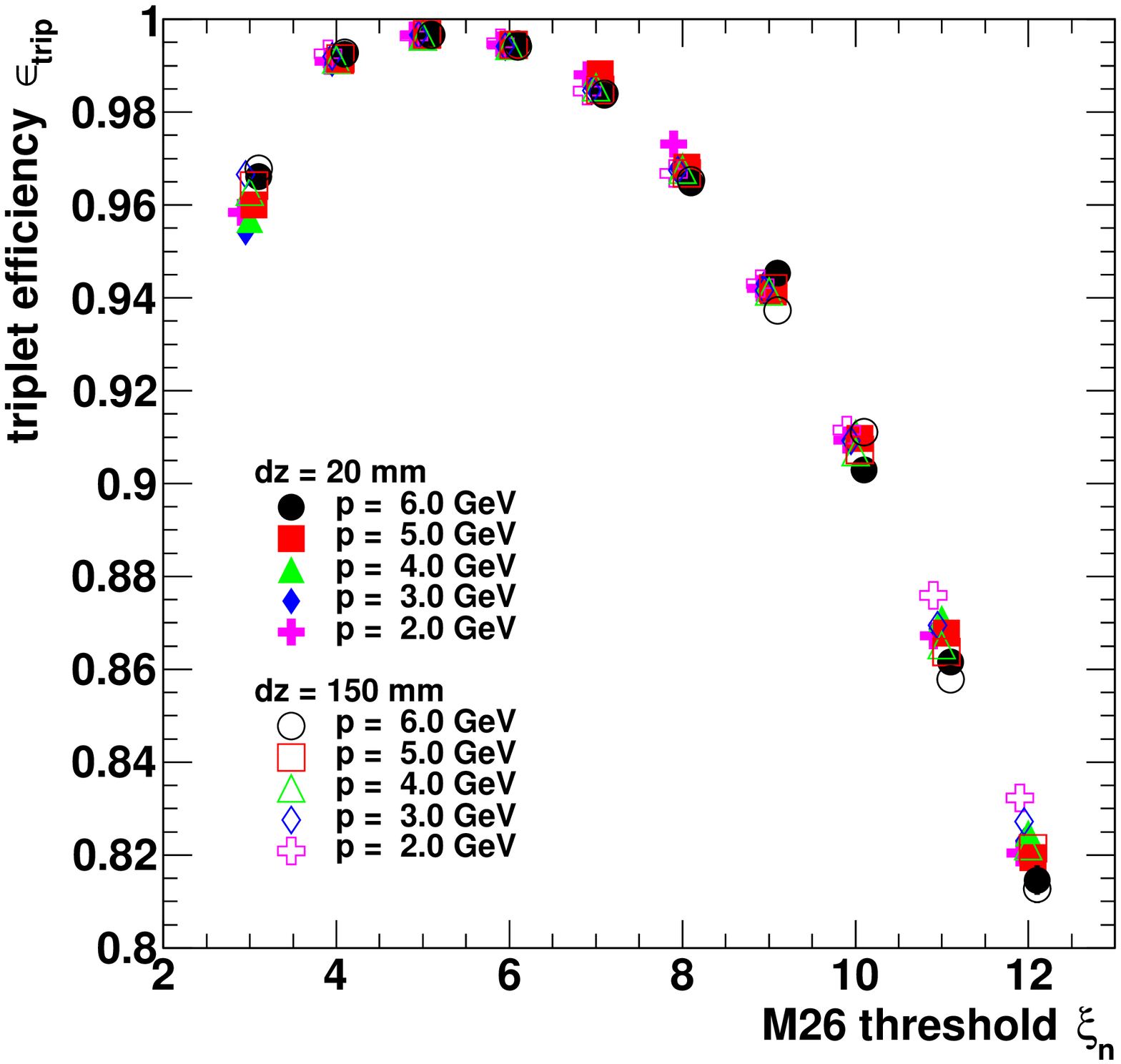}	\put( -30,158){(A)}
  \includegraphics[width=0.49\textwidth]{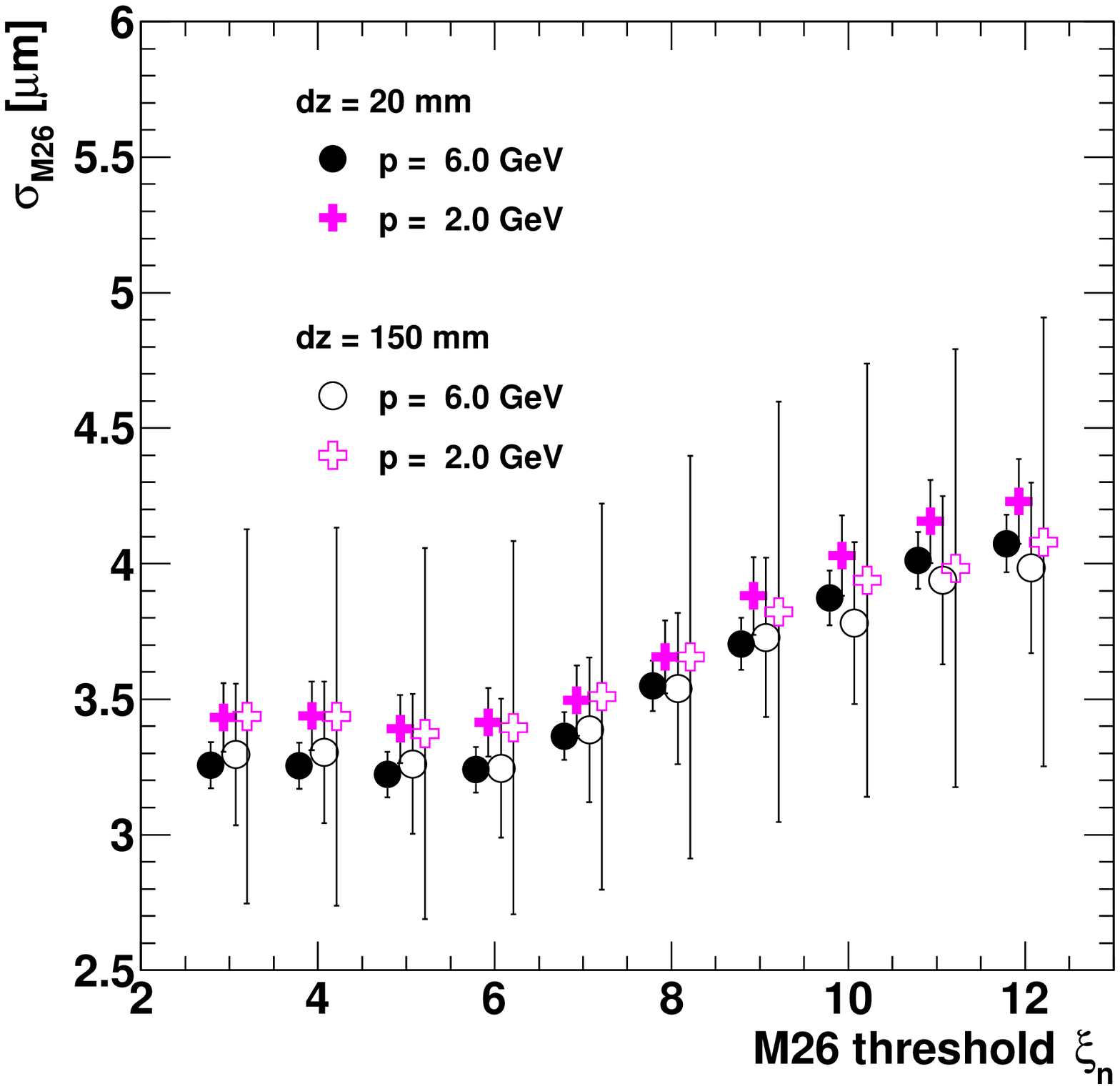}	\put( -30,158){(B)} 
  \else
  \includegraphics[width=0.49\textwidth]{effi_vs_thres.eps}	\put( -30,158){(A)}
  \includegraphics[width=0.49\textwidth]{reso_vs_thres.eps}	\put( -30,158){(B)} 
  \fi
  \caption[Telescope intrinsic sensor resolution for different threshold settings, beam energies and geometries~\cite{ref:thomas}]{
(A) Average efficiency of all telescope sensors for different beam energies and sensor spacing vs.~applied threshold.
(B) The measured intrinsic resolution of the $\sigma_{\textrm{M26}}$ for different beam energies $p$ and sensor spacing $\textrm{d}z$ as a function of the applied sensor threshold $\noise$.
In both images values are shifted on the $x$-axis for improved legibility.}
  \label{fig:resivsenergy_thresh}
\end{figure}

\noindent
A margin $d = 50\,\micro\meter$ is chosen for 6\,GeV and $\dz = 20\,\milli\meter$, which is scaled with the inverse energy and the plane distance $\dz$ between measurements. 
In figure~\ref{fig:resivsenergy_thresh}~(A) the telescope triplet efficiency dependence on the sensor threshold is shown for various beam energies and sensor spacings.
Statistical uncertainties are negligible.
Systematic uncertainties are insignificant due to a precise alignment and the sufficiently large margin $d$.  
For thresholds $\noise = 5$ and $\noise = 6$ the efficiency is measured to $99.6\,\%$ and $99.4\,\%$, respectively.
With increasing threshold, the efficiency decreases to about $82\,\%$ at $\noise = 12$. 
The efficiency varies by less than 0.1\,\% between different geometries and energy settings at thresholds $\noise = 5$ and $\noise = 6$. 

The threshold level affects not only the efficiency, but also the intrinsic resolution. 
With an increased threshold, a hit is formed on average from smaller clusters which deteriorates its position estimate. 
A deterioration also occurs towards lower thresholds, which allow for an increased number of noise induced signals to pass the zero-suppression on the chip.
Figure~\ref{fig:resivsenergy_thresh}~(B) shows the intrinsic sensor resolutions $\sigmai$ derived from the biased track fits
 for different beam energies and plane distances as a function of the applied sensor threshold.
An agreement of the various measurements is found within their uncertainties. 
The minimum of $\sigmai$ is reached for a sensor threshold setting of $\noise = 5$.
A previous measurement and analysis~\cite{ref:j.behrmeasurements, ref:joerg} taken at $\noise = 10$ yielded $\sigma_{\textrm{M26}} = (4.35\,\pm\,0.10)\,\micro\meter$,
 which seems to slightly overestimate the intrinsic resolution.

\subsection{Track resolution predictions using General Broken Lines}

%
Using the measured intrinsic resolution and the material budget in the beam (telescope planes, air, DUT),
 predictions of the track resolution at the actual DUT position $\zdut$ (cf.~figure~\ref{fig:datura_sketch}) are rendered possible.
Therefore, it is possible to perform an a priori calculation of the optimal telescope geometry for a certain measurement set-up~\cite{spannagel_2016_48795}. 
In this section, $\zdut$ is assumed to be in the centre between plane\,2 and plane\,3.


\begin{figure}[b!]
  \centering
  \ifdefined\notFOREPJ
  \includegraphics[width=0.49\textwidth]{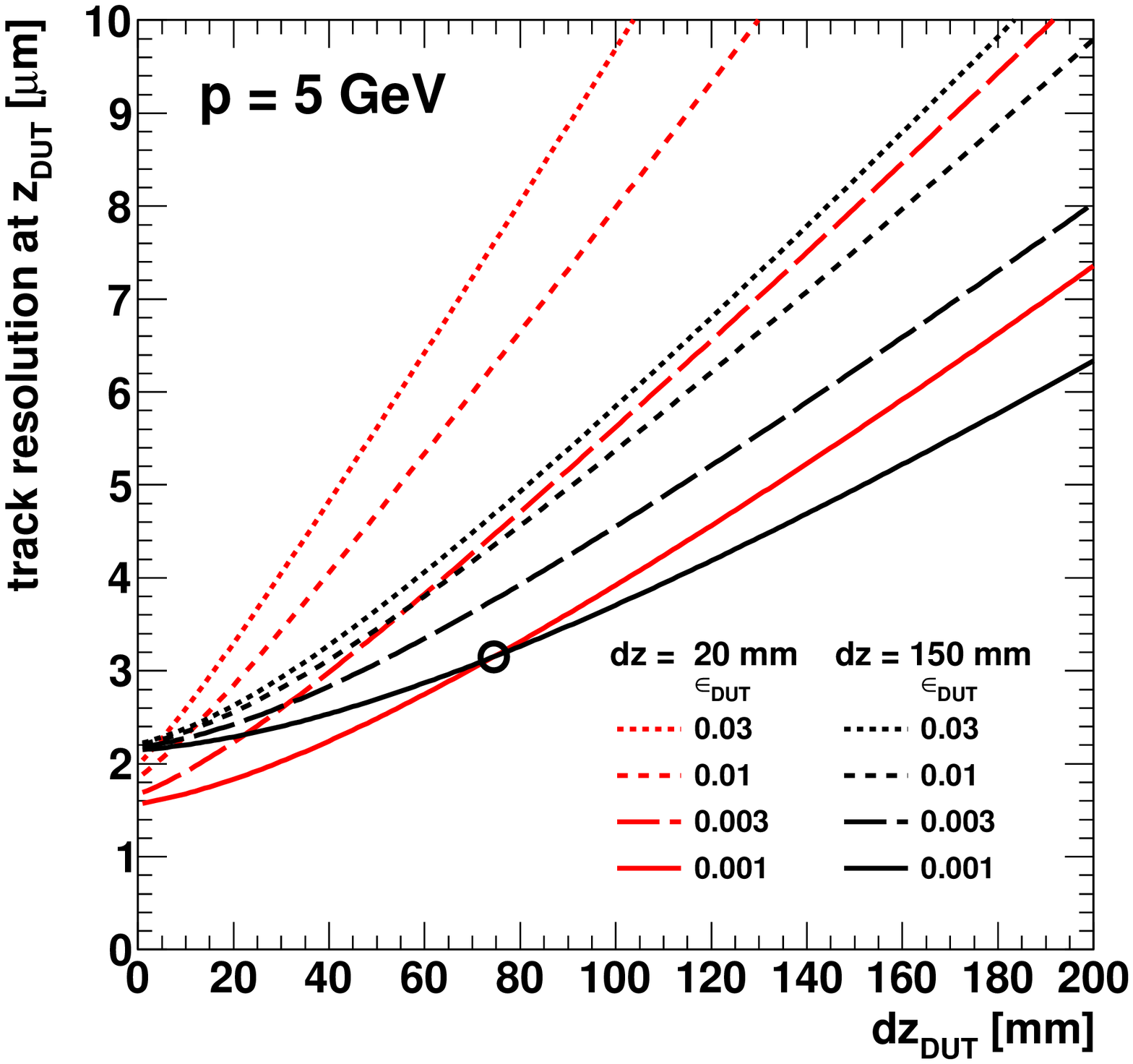}\put(-162,32){(A)}
  \includegraphics[width=0.49\textwidth]{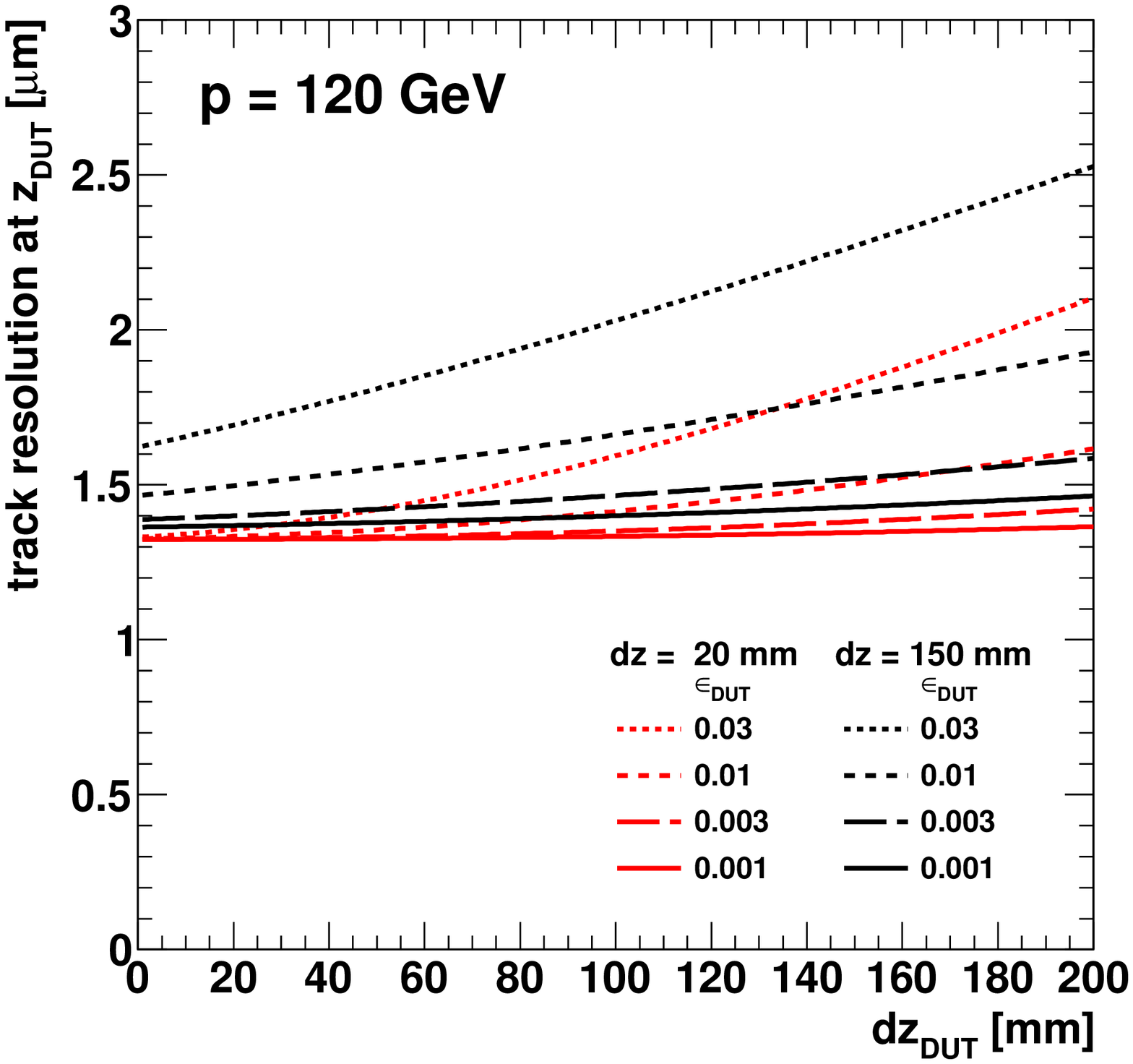} \put(-162,32){(B)}
  \else
  \includegraphics[width=0.49\textwidth]{trackres_vs_dzdut_DESY.eps}\put(-162,32){(A)}
  \includegraphics[width=0.49\textwidth]{trackres_vs_dzdut_SPS.eps} \put(-162,32){(B)}
  \fi
  \caption[Track resolution for various material budgets as a function of the distance between DUT and neighbouring planes]{
  The track resolution for various material budgets is shown as a function of the equidistant spacing between DUT and neighbouring planes at 5\,GeV (A), and 120\,GeV (B)
  at the centre of the beam telescope ($\zdut$).}
  \label{fig:CalcResos_dzdut}
\end{figure}

Using the GBL formalism, the track resolution is analytically calculated at points of interest along the particle trajectory. 
The track resolution at the DUT for four different DUT material budgets $\epsdut$ is depicted as a function of the spacing $\dzdut$ in figure~\ref{fig:CalcResos_dzdut}.
Exemplified are two configurations with $\dz = 20\,\milli\meter$ and $\dz = 150\,\milli\meter$ at (A) DESY-II and (B) SPS beam energies. 
For $\dz = \dzdut = 20\,\milli\meter$ at 5\,GeV beam energy the track resolution at the DUT is 
\begin{equation}
 \sigmat(\zdut) = \left( 1.83\,\pm\,0.03\right)\,\upmu\meter 
\end{equation}

\noindent
for $\epsdut = 0.001$.
For SPS energies, a track resolution of $(1.33\,\pm\,0.03)\,\upmu\meter$ is predicted.
The shown resolutions monotonically increase with increasing $\dzdut$. 
In order to achieve the best possible resolution, the inner $\Mimosa$ planes should therefore be positioned as close as possible to the DUT, i.e.~$\dzdut$ is to be minimised. 
In addition, figure~\ref{fig:CalcResos_dzdut}~(A) shows, that the optimal plane spacing $\dz$ depends on the actual DUT size along the beam direction and its material budget.
For instance for a DUT with a material budget of $\epsdut = 0.001$ (solid lines) at 5\,GeV, a narrow configuration shows a higher track resolution for $\dzdut < 74\,\milli\meter$,
 while a wide configuration is best for $\dzdut > 74\,\milli\meter$. 
The intersection is marked with a black circle in figure~\ref{fig:CalcResos_dzdut}~(A). 
The position of the intersection depends on the material budget of the DUT $\epsdut$ and shifts to smaller $\dzdut$ with increasing material budget. 
At SPS energies, the resolution functions for the wide and the narrow configuration do not intersect, with the narrow configuration showing a slightly better track resolution at the DUT. 
However, the difference is less than $0.5\,\micro\meter$ for $\zdut = 100\,\milli\meter$ even at $\epsdut = 0.03$. 

%
%

\begin{figure}[t]
  \centering
  \ifdefined\notFOREPJ
  \includegraphics[width=0.49\textwidth]{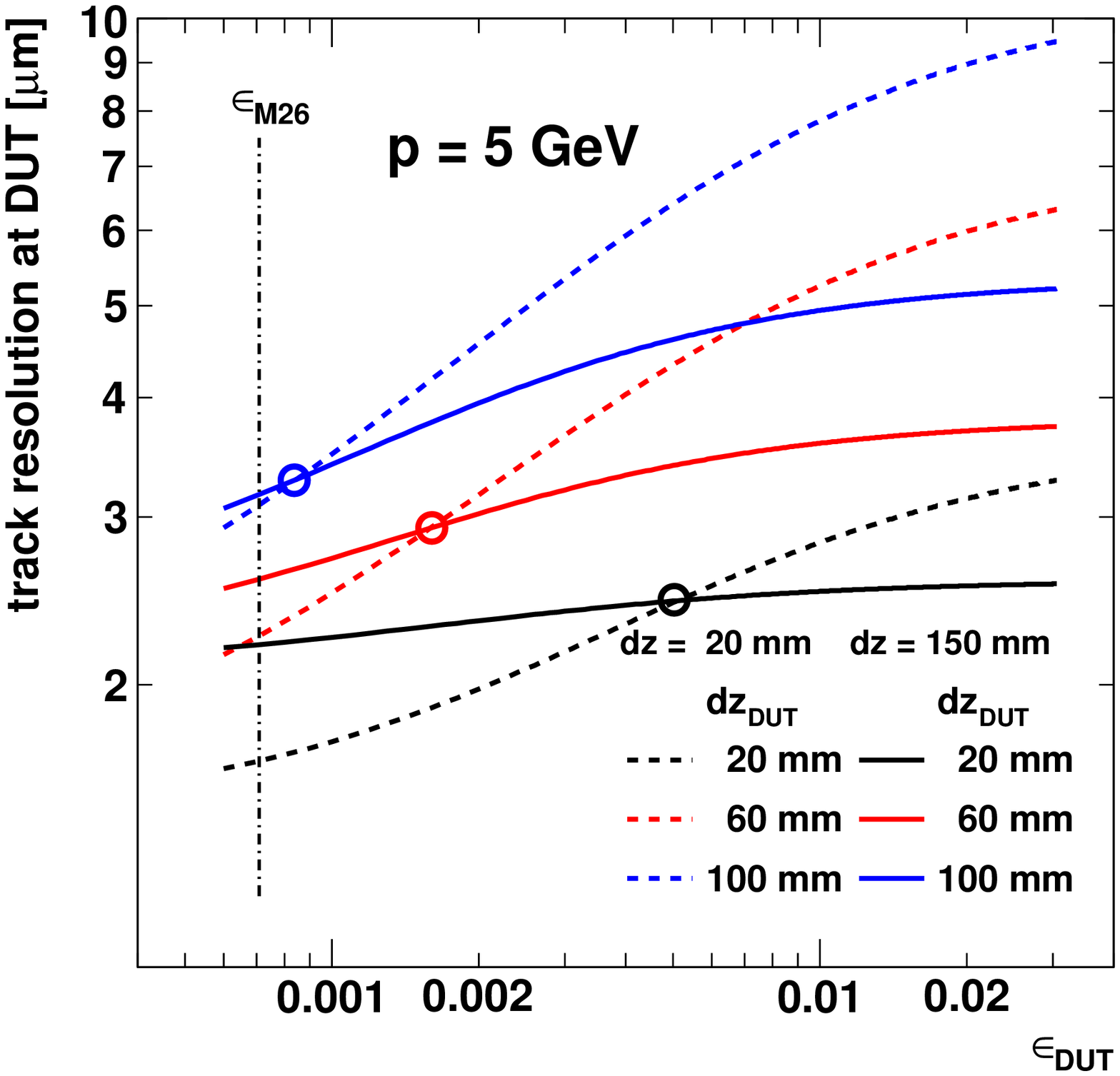} \put(-142,32){(A)}
  \includegraphics[width=0.49\textwidth]{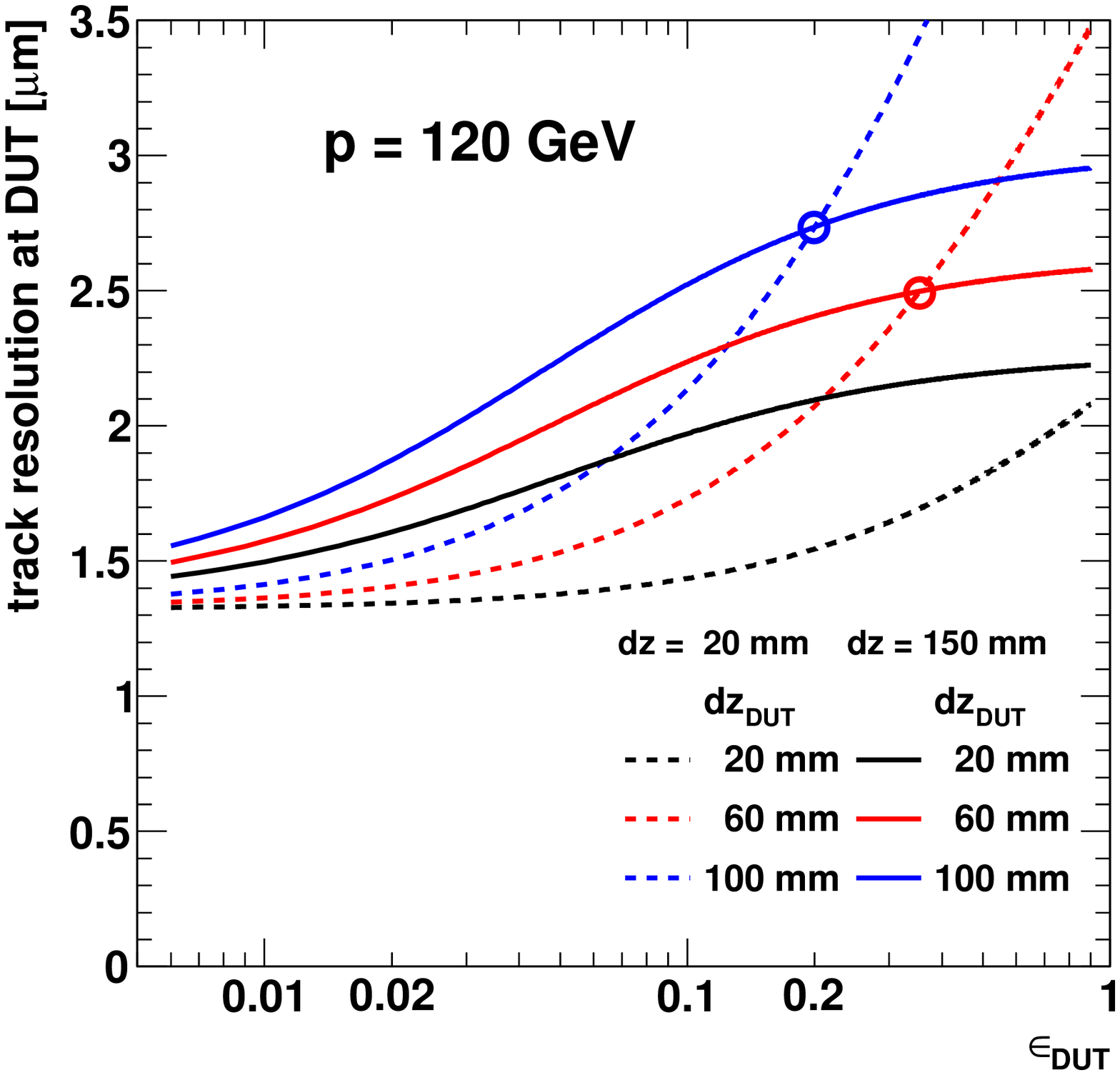} \put(-142,32){(B)}
  \else
  \includegraphics[width=0.49\textwidth]{trackres_vs_epsdut_DESY.eps} \put(-142,32){(A)}
  \includegraphics[width=0.49\textwidth]{trackres_vs_epsdut_SPS.eps} \put(-142,32){(B)}
  \fi
   \caption[Track resolution as a function of the beam energy]{
   The calculated track resolutions at the DUT for two geometries are shown as a function of $\epsdut$ at $\zdut$ at a beam energy of 5\,GeV (A) and 120\,GeV (B). 
   Note the double logarithmic scale for figure (A). 
   }
 \label{fig:CalcResoP_DUT}
\end{figure}

In figure~\ref{fig:CalcResoP_DUT} the achievable track resolution at $\zdut$ as a function of the material budget is shown for a beam energy of 5\,GeV (A) and 120\,GeV (B).
Dashed and solid lines represent calculations for $\dz = 20\,\milli\meter$ and $\dz = 150\,\milli\meter$, respectively. 
The track resolution deteriorates with increasing $\epsdut$. 
The optimal plane spacing $\dz$ depends on the actual size requirements for the DUT along the beam direction and its material budget.
For instance at 5\,GeV, for a DUT with a material budget of $\epsdut = 0.002$, a plane spacing of $\dz = 20\,\milli\meter$ should be used, if $\dzdut$ is as small as 20\,mm. 
However, a plane spacing of $\dz = 150\,\milli\meter$ is preferred, if $\dzdut$ is 60\,mm or larger. 
Intersections are marked with open circles. 
At 120\,GeV, the intersection for $\dzdut = 100\,\milli\meter$ occurs at a DUT material budget of $\epsdut = 0.2$, which is considerably larger compared to the case of DESY beam energies.
For $\epsdut$ below (above) the intersection, a narrow (wide) configuration is preferable. 
The position of the intersection shifts to smaller material budgets with increasing $\dzdut$. 

A web tool yielding compatible results in comparison with the GBL calculations is available~\cite{webtool}. 
The tool calculates track resolutions for a fixed set-up with six planes and one DUT in the centre. 
A more versatile GBL track resolution calculator allowing for all possible geometries is also available~\cite{gbltool}. 

%% file: content/dutintegration.tex
DUTs can be mechanically integrated into an $\eudet$-type beam telescope at three positions: upstream or downstream of the telescope, or between the two telescope arms.
If placed between the arms, micrometer precision $xy\phi$-stages are available for translation scans and rotation of the DUT~\cite{Mimosa-twiki}.
At this position, the maximum width of the DUT set-up is 500\,mm.
DUTs with larger spatial dimensions are therefore placed at the downstream end of the beam telescope. 

User DAQ systems are integrated to the TLU the same way as the beam telescope DAQ itself using either the RJ45 or the LEMO interface, cf.~section~\ref{sec:tdaq}.
The handshake mode is configurable for each integrated system individually. 
For the integration of the DUT data stream with EUDAQ, a producer capable of receiving commands by the Run Control and sending data to the $\eudaq$ Data Collector is necessary
 as described in section~\ref{sec:eudaq}.

A dedicated alignment run prior to data taking with no DUT in the beam allows for precise alignment of the telescope planes, especially for larger $\epsdut$. 
With a proper alignment at hand, runs including a DUT are to be analysed subsequently in order to align the DUT with respect to the beam telescope. 
The reconstructed tracks can then be used to characterise the DUT, i.e.~to measure its intrinsic resolution or efficiency. 

With increasing $\epsdut$, and thus multiple scattering within the DUT, the choice of the tracking algorithm needs further consideration. 
In general, a GBL fit produces tracks with a lower $\chi^2$ compared to straight line fits,
 as kinks at the possibly thick DUT and also at the $\Mimosa$ planes themselves are allowed for.
Therefore, using GBL for track fitting is recommended. 

For comparison, in a narrow configuration using a 5\,GeV electron/positron beam,
 the track resolution of a EUDET-type beam telescope at a DUT with $\epsdut = 0.001$ using GBL is about $3.7\,\upmu\meter$ making use of all six planes at $\dz = 150\,\milli\meter$ and $\dzdut = 100\,\milli\meter$.
The optimal configuration using only the upstream planes is a wide configuration with $\dz = 150\,\milli\meter$,
 and the achievable track resolution is about $8.2\,\upmu\meter$. 
Hence, in this case, using only the upstream planes deteriorates the resolution by approximately $4.5\,\upmu\meter$.

%% file: content/conclusion.tex

In this work the $\eudet$-type beam telescopes have been described and their performance has been investigated. 
The highly flexible and versatile $\eudet$-type beam telescopes come with a Trigger Logic Unit with time stamping capabilities and a clearly defined interface for an integration of user data acquisition systems.
It is complemented by a data acquisition system for the telescope sensors, and the powerful software packages $\eudaq$ and $\EUTelescope$
 --  a modular data acquisition framework and an offline reconstruction software, respectively. 
The intrinsic resolution of the $\Mimosa$ sensors have been measured to be $(3.24 \pm 0.09)\,\upmu\meter$.
General Broken Lines calculations predict a track resolution of $(1.83 \pm 0.03)\,\upmu\meter$ on the DUT with 20\,mm plane spacing for thin sensors with $\epsdut = 0.001$ at 5\,GeV. 

Continuous development efforts are ongoing to upgrade and enhance the Trigger Logic Unit, the data acquisition system, and offline reconstruction framework~\cite{ref:tipp2014_eudaq}.  
The maximum achievable rate will be increased by the successor of the TLU. 
A new plane comprising four MIMOSA\,28 planes has been built for the AIDA telescope, covering a detection area of $16\,\centi\meter^2$, and an appropriate DAQ has been developed to cope with the higher data rates. 
Future versions of \eudaq\ will provide support for asynchronous data streams allowing to operate certain devices at a much higher trigger frequency than others
 and thus making full use of the the beams available in the various beam lines.
The development of the $\EUTelescope$ analysis software is driven by the large and diverse user community. 
Features currently under development include tracking in magnetic fields and more accurate detector geometry descriptions. 